\documentclass[iop,apj,tighten]{emulateapj}
\usepackage[breaklinks,colorlinks,urlcolor=blue,citecolor=blue,linkcolor=blue]{hyperref}

\usepackage{graphicx}
\usepackage{bm}
\usepackage{amsmath,amssymb}
\usepackage{cleveref}
\usepackage{appendix}
\usepackage{color}
\newcommand{\Msun}{{\rm M}_\odot}
\usepackage{paralist}

\shorttitle{Spins and Masses for Hierarchical Mergers}
\shortauthors{Tagawa et al.}

\begin{document}
\title{Signatures of Hierarchical Mergers in Black Hole Spin and Mass distribution}

\author{
Hiromichi Tagawa\altaffilmark{1}, 
Zolt{\'a}n Haiman\altaffilmark{2}, 
Imre Bartos\altaffilmark{3}, 
Bence Kocsis\altaffilmark{4}, 
Kazuyuki Omukai\altaffilmark{1}
}
\affil{\altaffilmark{1}Astronomical Institute, Graduate School of Science, Tohoku University, Aoba, Sendai 980-8578, Japan\\
\altaffilmark{2}Department of Astronomy, Columbia University, 550 W. 120th St., New York, NY, 10027, USA\\
\altaffilmark{3}Department of Physics, University of Florida, PO Box 118440, Gainesville, FL 32611, USA\\
\altaffilmark{4}Department of Physics, Oxford University
}
\email{E-mail: htagawa@astr.tohoku.ac.jp}

\begin{abstract} 
Recent gravitational wave (GW) observations by LIGO/Virgo show evidence for hierarchical mergers, where the merging BHs are the remnants of previous BH merger events. These events may carry important clues about the astrophysical host environments of the GW sources.
In this paper, we present the distributions of the effective spin parameter ($\chi_\mathrm{eff}$), the precession spin parameter ($\chi_\mathrm{p}$), and the chirp mass ($m_\mathrm{chirp}$) expected in hierarchical mergers. 
Under a wide range of assumptions, hierarchical mergers produce
(i) a monotonic increase of the average of the typical total spin for merging binaries, which we characterize with ${\bar \chi}_\mathrm{typ}\equiv \overline{(\chi_\mathrm{eff}^2+\chi_\mathrm{p}^2)^{1/2}}$, up to roughly the maximum $m_\mathrm{chirp}$ among first-generation (1g) BHs, 
and (ii) a plateau at ${\bar \chi}_\mathrm{typ}\sim 0.6$ at higher $m_\mathrm{chirp}$. 
We suggest that the maximum mass and typical spin magnitudes for 
1g BHs can be estimated 
from ${\bar \chi}_\mathrm{typ}$ as a function of $m_\mathrm{chirp}$. 
The GW data observed in LIGO/Virgo O1--O3a prefers an increase in ${\bar \chi}_\mathrm{typ}$ at low $m_\mathrm{chirp}$, 
which is consistent with the growth of the BH spin magnitude by hierarchical mergers, at $\sim 2 \sigma$ confidence. 
A Bayesian analysis using the $\chi_\mathrm{eff}$, $\chi_\mathrm{p}$, and $m_\mathrm{chirp}$ distributions 
suggests that 1g BHs 
have the maximum mass of $\sim 15$--$30\,\Msun$ if the majority of mergers are of high-generation BHs (not among 1g-1g BHs), 
which is consistent with mergers in active galactic nucleus disks and/or nuclear star clusters, 
while if mergers mainly originate from globular clusters, 1g BHs are favored to have non-zero spin magnitudes of $\sim 0.3$. 
We also forecast that signatures for hierarchical mergers in 
the ${\bar \chi}_\mathrm{typ}$ distribution can be confidently recovered once the number of GW events increases to $\gtrsim O(100)$. 
\end{abstract}

\section{Introduction}

Recent detections of gravitational waves (GWs) by LIGO \citep{2015CQGra..32g4001L} and Virgo \citep{2015CQGra..32b4001A} have shown evidence for a high rate of black hole (BH)-BH and neutron star (NS)-NS mergers in the Universe
\citep{TheLIGO18,Venumadhav19,LIGO20_O3_Catalog}. 
However, proposed astrophysical pathways to mergers remain highly
debated.  Indeed there are currently 
a large number of
such possible pathways, with widely different environments and
physical processes.  A possible list of these currently includes 
isolated binary evolution \citep[e.g.][]{Dominik12,Kinugawa14,Belczynski16,Spera19} 
accompanied by mass transfer \citep[][]{Pavlovskii17,Inayoshi17,vandenHeuvel17}, common envelope ejection \citep[e.g.][]{Paczynski76,Ivanova13}, envelope expansion \citep{Tagawa18}, chemically homogeneous evolution in a tidally distorted binary  \citep{deMink16,Mandel16,Marchant16}, 
evolution of triple or quadruple systems
\citep[e.g.][]{Silsbee17,Antonini17,Michaely19,Fragione19}, 
gravitational capture 
\citep[e.g.][]{OLeary09,Gondan18a,Rasskazov19}, 
dynamical evolution in open clusters
\citep[e.g.][]{Banerjee17,Kumamoto18,Rastello18}, 
young stellar clusters
\citep[e.g.][]{Ziosi14,DiCarlo19,Rastello20}, 
and dense star clusters 
\citep[e.g.][]{PortegiesZwart00,Samsing14,OLeary16,Rodriguez16,Fragione18_EvolvingGC,Fragione19b_GN}, 
and interaction in active phases of galactic nucleus (AGN) disks
\citep[e.g.][]{Bartos17,Stone17,McKernan17,Tagawa19}. 

Recently several GW events were reported by LIGO and Virgo whose measured physical properties pose interesting constraints on their astrophysical origin. 
These include nine candidates for mergers in the upper-mass gap ($\sim 50$--$130\,\Msun$) such as GW190521  \citep{TheLIGO18,Zackay19_GW170817A,LIGO20_O3_Catalog,LIGO20_GW190521,LIGO20_GW190521_astro}. 
Additionally, mergers with very unequal masses have been reported -- GW190412 ($q=0.28^{+0.13}_{-0.07}$, \citealt{LIGO20_GW190412}) and GW190814 ($q=0.112^{+0.008}_{-0.009}$, \citealt{LIGO20_GW190814}) --, which are also atypical in stellar evolutionary models of isolated binaries
\citep{Gerosa20,Olejak20_GW190412,Zevin2020_GW190814}. 
The object in the lower mass gap in GW190814 and a non-zero spin for the primary BH ($a_1 = 0.43^{+0.16}_{-0.26}$) in GW190412 are consistent with a scenario in which the merging compact objects (COs) had experienced  previous episode(s) of mergers or significant accretion. 
These events suggest that growth by gas accretion or hierarchical mergers may be common among COs \citep[see e.g.][]{OLeary16,Gerosa20,LIGO20_GW190521_astro,Hamers20_GW190412,Rodriguez20_GW190412,Safarzadeh20_GW190814,Safarzadeh20_GW190521,Yang20_GW190814,LiuLai20_GWevents,Tagawa20_ecc,Tagawa20_massgap,Samsing20,Fragione2020_Retantion}. 


Hierarchical mergers may occur in dynamical environments, such as globular clusters (GCs), nuclear star clusters (NSCs), and active galactic nucleus (AGN) accretion disks. 
In GCs, up to $\sim10-20\%$ of detected mergers may be caused by high-generation (high-g) BHs depending on spin magnitudes of 1g BHs \citep{OLeary16,Rodriguez19_hierarchical}. 
Repeated mergers of BHs and stars may produce intermediate-mass BHs (IMBHs, BHs with masses of $\sim 100$--$10^4\,\Msun$) in NSCs without supermassive BHs \citep[SMBHs,][]{Antonini19,Askar20,Mapelli20}. 
In NSCs with SMBHs, it is uncertain how often hierarchical mergers occur \citep[e.g.][]{ArcaSedda20}. 

In AGN disks, hierarchical mergers are predicted to be frequent due to the high escape velocity and efficient binary formation and evolution facilitated by gaseous \citep{Yang19b_PRL,McKernan19} and stellar interactions \citep{Tagawa19}. 
\citet{Yang19b_PRL} and \citet{McKernan19,McKernan20_NSWD} 
identified the expected mass ratio and spin distribution of
hierarchical mergers in hypothetical migration traps (MTs) of AGN disks, defined to be regions where objects accumulate rapidly as they interact with the accretion disks analogously to planetary migration.\footnote{Note that the orbital radii where this takes place were derived by assuming Type-I migration \citep{Bellovary16}, but these assumptions may be inconsistent for BHs embedded in AGN disks as gaps may be opened in the accretion disks \citep[see e.g. Eqs.~45--46][]{Kocsis11}. Also, \citet[][]{Pan2021} found that 
the traps can disappear if radiation pressure is correctly accounted for.} 
\citet{Tagawa20b_spin,Tagawa20_massgap} showed that hierarchical mergers take place in AGN disks without MTs 
and derived the corresponding mass and spin distributions self-consistently. 
In the latter models \citep[e.g.][]{Tagawa20b_spin}, the mass and spin distributions of merging BHs are significantly different compared to those in the former models. This is mainly due to  binary-single interactions which take place frequently at large orbital radii where the gas density is very low and gas effects drive the binaries toward merger more slowly and allow ample time for such binary-single interactions.

Several authors have investigated the properties of GWs associated with hierarchical mergers 
\citep{Gerosa17,Yang19b_PRL,Kimball20,Doctor20}. 
\citet{Gerosa17} estimated the fraction of future detected sources contributed by hierarchical mergers under the assumption that first-generation (1g) BHs have a flat spin distribution and binary components are drawn independently.
\citet{Fishbach17_hierarchical} estimated the required number of events to detect hierarchical mergers using the distribution of the BH spin magnitudes. 
\citet{Doctor20} constructed a toy model to obtain the properties of hierarchical mergers from the distribution of subpopulations for BHs 
under various assumptions for coagulation and depletion in the population and constrained parameters using LIGO/Virgo O1--O2 data. 
\citet{Kimball20} examined whether the observed events in the same catalog are compatible with hierarchical mergers particularly in GCs. These models found no evidence for a high rate of hierarchical mergers in this early catalog.
More recently, by analyzing the ensemble of events detected during LIGO/Virgo's O1-O3a observing runs, 
\citet{Kimball2020b} and \citet{Tiwari20} found preference 
for at least one, but probably multiple hierarchical mergers in the detected sample. 
The conclusion of \citet{Kimball2020b} strongly depends on the assumed escape velocity in the host environment, with higher escape velocities favoring a larger number of hierarchical mergers.

In this paper, 
we focus on distributions of the effective and precession spin parameters ($\chi_\mathrm{eff}$ and $\chi_\mathrm{p}$) and the chirp mass ($m_\mathrm{chirp}$), and predict characteristic features in them expected from hierarchical mergers. 
We use $m_\mathrm{chirp}$ as this variable is most precisely determined by GW observations, and $\chi_\mathrm{eff}$ and $\chi_\mathrm{p}$ as these characterize the BH spin magnitudes in a binary. 
Here, $\chi_\mathrm{eff}$ and $\chi_\mathrm{p}$ are defined as 
\begin{equation}
\chi_\mathrm{eff}=\frac{m_1{a}_1\mathrm{cos}\theta_1+m_2{a}_2 \mathrm{cos}\theta_2}{m_1+m_2}
\end{equation}
and 
\begin{equation}
\chi_\mathrm{p}=\mathrm{max}\left({a}_1 \mathrm{sin}\theta_1, q\frac{4q+3}{4+3q}{a}_2 \mathrm{sin}\theta_2\right)
\end{equation}
\citep{Hannam2014,Schmidt2015}, 
where $m_1$ and $m_2$ are the masses, ${a}_1$ and ${a}_2$ are the spin magnitudes, $\theta_1$ and $\theta_2$ are the angles between the orbital angular momentum directions and the BH spins of the binary components, 
$q\equiv m_2/m_1\leq1$ is the mass ratio, and 
$m_\mathrm{chirp}\equiv\left(m_1m_2\right)^{3/5}\left(m_1 +m_2\right)^{-1/5}$. 
We identify and characterize features expected in hierarchical mergers using mock GW data, and 
find that intrinsic properties (maximum mass and typical spin magnitude) of 1g BHs can be constrained by recovering the features, which enables us to distinguish astrophysical models. 
By analyzing the GW data obtained in LIGO/Virgo O1--O3a, 
we investigate whether such features are consistent with observed GW data, 
and identify the astrophysical population models most consistent with the data.
Finally, using mock GW data, we estimate how well parameters characterizing the spin distribution 
can be recovered in future catalogs depending on the number of events.

The paper is organized as follows. 
In $\S$~2, we describe our method to construct mock GW data and detect signatures for hierarchical mergers. 
We present our main results in $\S$~3, and give our conclusions in $\S$~4.

\section{Method}

\subsection{Overview}

We introduce a mock dataset generated by a simple $N$-body toy model ($\S\,\ref{section:construction_gw_data}$),
which allows us to explore hierarchical mergers more generically ($\S\,\ref{sec:result_profiles_mock}$). 
To identify features in the distributions representative of  
hierarchical mergers, 
we use a simple analytic model characterizing the spin distribution profile ($\S\,\ref{sec:model_xeff}$), and apply it to the observed GW data ($\S\,\ref{sec:result_rec_x_obs}$) and the $N$-body toy model ($\S\,\ref{sec:result_rec_x}$). 
Furthermore, to assess how well model predictions match the observed GW data, 
we also use a Bayes factor to assess relative likelihoods of models (including the $N$-body toy model and a physical model for mergers in AGN disks adopted from our simulations in \citealt{Tagawa20_massgap}, $\S\,\ref{sec:results_bayesian_factor}$).

In the analyses, we mostly use ${\chi}_\mathrm{typ}\equiv(\chi_\mathrm{eff}^2+\chi_\mathrm{p}^2)^{1/2}$ as it characterizes the spin magnitudes of BHs in binaries, and  
it is easily calculated from the quantities $\chi_\mathrm{eff}$ and $\chi_\mathrm{p}$ taken from \citet{LIGO20_O12_HP} and \citet{LIGO21_O3_HP}. 
However one should be aware of the following properties of ${\chi}_{\rm typ}$. 
First, unlike $\chi_{\rm eff}$, $\chi_{\rm p}$ is not conserved up to 2PN \citep[e.g.][]{Gerosa20_xp}, suffering additional uncertainties due to its modulation. 
Second, due to the geometry, the contribution of $\chi_{\rm p}$ is on average larger than $\chi_{\rm eff}$ by a factor of $\sim 3$ in cases of isotropic BH spins (Eq.~\ref{eq:xtyp_xeff_xp} in the Appendix). 
Third, $\chi_{\rm p}$ is often unconstrained in the LIGO/Virgo events (e.g. Fig.~\ref{fig_app:mc_x_obs}).

\subsection{Constructing mock GW data}

\label{section:construction_gw_data}

To understand and analyze the distributions of $\chi_\mathrm{eff}$, $\chi_\mathrm{p}$, and $m_\mathrm{chirp}$ typically expected in hierarchical mergers, 
we employ mock GW data. 

\subsubsection{Overall procedure}
\label{section:procedue_gw_data}

We construct mock data by following the methodology of \citet{Doctor20}:

\begin{enumerate}[1]
\item 
Sample $N_\mathrm{1g}$ BHs from 1g population as described in $\S\,\ref{sec:1gBHs}$. We set $N_\mathrm{1g}=10^6$ to ensure a sufficient number for detectable mergers. We call this sample $S$. 

\item 
Choose $\omega N_\mathrm{ng}$ pairs from $S$ by weighing the pairing probability $\Gamma$ ($\S\,\ref{sec:1gBHs}$), 
where $\omega$ is the fraction of BHs that merge at each step, and $N_\mathrm{ng}$ is the number of BHs in the sample $S$ ($N_\mathrm{ng}=N_\mathrm{1g}$ in the first iteration). 

\item 
Compute the remnant mass and spin, and the kick velocity  for merging pairs assuming random directions for BH spins, where we use the method described in \citet{Tagawa20b_spin}. 
Update the sample $S$ by removing BHs that have merged, and adding merger remnants if the kick velocity is smaller than the escape velocity ($v_\mathrm{esc}$). 

\item 
Repeat steps 2-3 for $N_\mathrm{s}$ steps. 

\item 
Determine the fraction of detectable mergers by assessing whether signal-to-noise ratio (SNR) of mergers exceeds the detection criteria ($\S\,\ref{sec:mock_data}$). 
Randomly choose $N_\mathrm{obs}$ observed mergers from the detectable merging pairs. Add observational errors following $\S\,\ref{sec:mock_data}$, and construct a mock GW dataset. 
\end{enumerate}

By changing the underlying parameters of the merging binaries in mock GW data (${\bm \lambda}_0$; presented in the next section), we can construct various $\chi_\mathrm{eff}$, $\chi_\mathrm{p}$ and $m_\mathrm{chirp}$ distributions expected in hierarchical mergers. 
For example, $N_\mathrm{s}$ and $\omega$ 
influences the fraction of hierarchical mergers ($\propto \sim \omega^{N_\mathrm{s}}$), while $N_\mathrm{s}$ specifies the maximum generation and mass of BHs.

\subsubsection{First generation BHs and pairing}
\label{sec:1gBHs}

We assume that the masses of 1g BHs are drawn from the power-law distribution as 
\begin{align}\label{eq:p_mp}
p_{m_\mathrm{1g}}\propto
\left\{
\begin{array}{l}
m_\mathrm{1g}^{-\alpha}
\qquad~~~~~~\mathrm{for}~m_\mathrm{min}<m_\mathrm{1g}<m_\mathrm{max}, \\
0 
\qquad~~~~~~~~~~\mathrm{otherwise}, 
\end{array}
\right.
\end{align}
where $\alpha$ is the power-law slope, $m_\mathrm{min}$ and $m_\mathrm{max}$ are the minimum and maximum masses.

We set the dimensionless spin magnitude for 1g BHs to 
\begin{align}
\label{eq:a_ini}
   a_\mathrm{ini}=|a_\mathrm{ave} +a_\mathrm{uni} U[-1:1]|
\end{align}
where $U[-1:1]$ represent uniform distribution randomly chosen from -1 to 1, and $a_\mathrm{ave}$ and $a_\mathrm{uni}$ are parameters characterising initial spins of 1g BHs. 
We assume that the spin magnitude for 1g BHs does not depend on the masses of 1g BHs. 
This assumption may be justified for single BHs, for which slow rotation is motivated by theoretical considerations \citep{Fuller19_massive}. 
Here we assume $a_\mathrm{ave}=a_\mathrm{uni}=0$ in the fiducial model. 
On the other hand, for mergers of field binaries (FBs), a large fraction of BHs may experience tidal synchronization, and the dispersion of the BH spin magnitudes decreases with BH masses \citep[e.g.][]{Hotokezaka17,Bavera19,Safarzadeh20}. The spin distribution expected in this pathway is considered in $\S\,$\ref{sec:results_several_pop}. 

We assume the redshift distribution of merging BHs as 
\begin{align}
\label{eq:p_z}
   p_{z}\propto &\frac{dV_c}{dz}\frac{1}{1+z}
\end{align} 
so that a merger rate density is uniform 
in comoving volume and source-frame time. 
Here, $dV_c/dz$ is calculated assuming $\Lambda$CDM cosmology with the Hubble constant $H_0 \simeq 70\,\mathrm{km/s/Mpc}$, the matter density today $\Omega_\mathrm{m0}=0.24$, and the cosmological constant today $\Omega _{\Lambda0}=0.76$ \citep{Plank16}. 
We also investigate different choices in $\S\,\ref{sec_app:result_pop_parameters}$ (see also \citealt{Fishbach18,Yang20}). 
We set the maximum redshift to be $1.5$ considering LIGO/Virgo sensitivities \citep{LIGO19_IMBH}.

To draw merging pairs, 
we simply assume that the interaction rate depends on the binary masses 
with a form 
\begin{align}
\label{eq:p_zmpq}
  \Gamma
  \propto \left( m_\mathrm{1}+m_\mathrm{2}\right)^{\gamma_\mathrm{t}} q^{\gamma_\mathrm{q}}
\end{align}
as employed in \citet{Doctor20}. 
This parameterization enables us to mimic the effects that massive and equal-mass binaries are easy to merge in plausible models due to exchanges at binary-single interactions, mass segregation in clusters, interaction with ambient gas, mass transfer, or common-envelope evolution \citep[e.g.][]{OLeary16,Rodriguez19_hierarchical,Tagawa20_massgap,Olejak20_GW190412}. 

Using the model described above and adding observational errors ($\S\,$\ref{sec:mock_data}), we can construct a mock observational dataset. 
The parameter set characterizing a mock dataset is 
$\bm \lambda_0 = \{\alpha, m_\mathrm{min},m_\mathrm{max}, a_\mathrm{ave}, a_\mathrm{uni},\gamma_\mathrm{t},\gamma_\mathrm{q},\omega,N_\mathrm{s}, v_\mathrm{esc}, N_\mathrm{obs} \}$. 
The fiducial choice of $\bm \lambda_0$ is described in 
$\S\,\ref{sec:numerical_choice}$ and Table~\ref{table:parameter_fiducial}.

\subsubsection{Mock observational errors}

\label{sec:mock_data}

To construct mock GW data, we need to put observational errors on observables. 
The true values of observables ${\bm \theta}$ are produced through the procedures in $\S\,\ref{section:procedue_gw_data}$ and $\S\,\ref{sec:1gBHs}$ assuming a set of the population parameters ${\bm \lambda}_0$. 
To incorporate observational errors to the mock data, we refer to the prescription in \citet{Fishbach20}. 
We assume that the binary is detected if the SNR of the signal in a single detector exceeds 8. 
We set the typical SNR, $\rho_0$, of a binary with parameters $m_\mathrm{chirp}$, $\chi_\mathrm{eff}$, and the luminosity distance $d_\mathrm{L}$ to 
\begin{align}
\label{eq:rho0}
\rho_{0} = 8 \left[\frac{m_\mathrm{chirp}(1+z)}{m_\mathrm{chirp,8}}\right]^{5/6}
\frac{d_\mathrm{L,8}}{d_\mathrm{L}} \left( 1+ \frac{3}{8}\chi_\mathrm{eff}\right)
\end{align}
where we fix $m_\mathrm{chirp,8}=10\,\Msun$ and $d_\mathrm{L,8}=1\,\mathrm{Gpc}$~(see eq.~26 in \citealt{Fishbach18}).
This scaling approximates the amplitude of a GW signal,  $m_\mathrm{chirp,8}$ and $d_\mathrm{L,8}$ are chosen to roughly match the typical values detected by LIGO at design sensitivity \citep{Chen17}, and the dependence on $\chi_\mathrm{eff}$ roughly reproduces results in \citet{LIGO19_IMBH}. 
We calculate $d_\mathrm{L}$ from $z$ assuming $\Lambda$CDM cosmology as stated above. 
The true SNR depends on the angular factor $\Theta$, 
and is given by 
 \begin{align}
\label{eq:rho_rho0}
\rho = \rho_0 \Theta . 
\end{align}
$\Theta$ plays the combined role of the sky location,
inclination, and polarization on the measured GW amplitude.
We tune the width of the  distribution to control the
uncertainty of the measured signal strength, which in turn
controls the uncertainty on the measured luminosity distance. 
We simply set $\Theta$ to a log-normal distribution with 
 \begin{align}
\label{eq:theta_dist}
\mathrm{log} \Theta \sim N\left(0,\frac{0.3}{1+\frac{\rho_0}{8}}\right) 
\end{align}
following \citet{Fishbach18}. 

From the true parameters $\rho$, $m_\mathrm{chirp}(1+z)$, $z$, 
$\chi_\mathrm{eff}$ and $\Theta$, 
we assume that the four parameters, 
the SNR ($\rho_\mathrm{obs}$), 
the chirp mass ($m_\mathrm{chirp,obs}$), 
$\chi_\mathrm{eff,obs}$, and 
$\chi_\mathrm{p,obs}$, are given with errors as below. 
We assume that the fractional uncertainty on the detector-frame chirp mass is 
\begin{align}\label{eq:sig_mchirp}
\sigma_{m_\mathrm{chirp}} =\frac{8}{\rho_\mathrm{obs}} 0.04\, m_\mathrm{chirp} (1+z), 
\end{align}
that on the SNR is 
 \begin{align}
\label{eq:sig_sn}
\sigma_{\rho} = 1
\end{align}
following \citet{Fishbach20}, 
and that on $\chi_\mathrm{eff}$ and $\chi_\mathrm{p}$ is, respectively,  
 \begin{align}
\label{eq:sig_xeff}
\sigma_{\chi_\mathrm{eff}} = 0.1\frac{8}{\rho_\mathrm{0}}, 
\end{align}
and 
\begin{align}
\label{eq:sig_xp}
\sigma_{\chi_\mathrm{p}} = 0.2\frac{8}{\rho_\mathrm{0}}, 
\end{align}
which roughly match typical observational error magnitudes in \citet{TheLIGO18} and \citet{LIGO20_O3_Catalog}. 
We assume that the observed median values ${\Tilde m}_\mathrm{chirp,obs}$, ${\Tilde \rho}_\mathrm{obs}$, ${\Tilde \chi}_\mathrm{eff,obs}$, and ${\Tilde \chi}_\mathrm{p,obs}$, respectively, 
from a normal distribution centered on the true values 
$m_\mathrm{chirp}(1+z)$, 
$\rho$, $\chi_\mathrm{eff}$, and $\chi_\mathrm{p}$ 
with the standard deviation $\sigma_{m_\mathrm{chirp}}$, $\sigma_{\rho}$, $\sigma_{\chi_\mathrm{eff}}$, and $\sigma_{\chi_\mathrm{p}}$. 
We further assume that 
the posterior distributions of ${m}_\mathrm{chirp}$, ${\rho}$, ${\chi}_\mathrm{eff}$, and ${\chi}_\mathrm{p}$ 
including errors 
for GW data in the $i^{\rm th}$ event 
are, respectively, calculated by drawing 
from a normal distribution centered on ${\Tilde m}_\mathrm{chirp,obs}$, ${\Tilde \rho}_\mathrm{obs}$, ${\Tilde \chi}_\mathrm{eff,obs}$, and ${\Tilde \chi}_\mathrm{p,obs}$ 
with the standard deviation $\sigma_{m_\mathrm{chirp}}$, $\sigma_{\rho}$, $\sigma_{\chi_\mathrm{eff}}$, and $\sigma_{\chi_\mathrm{p}}$. 
An observed value of $z$ is calculated from $d_\mathrm{L}$ derived by incorporating the observed values to Eq.~\eqref{eq:rho0} and the relation between $z$ and $d_\mathrm{L}$ so that Eq.~\eqref{eq:rho0} is valid for derived $z$.

\subsubsection{Numerical choices}
\label{sec:numerical_choice}

Table~\ref{table:parameter_fiducial} lists the parameter values adopted in the fiducial model. 
Referring to \citet{Fuller19_massive}, 
we set small BH spin magnitudes for 1g BHs as $a_\mathrm{ave}=a_\mathrm{uni}=0$. 
The power-law slope in the mass function for 1g BHs is given as $\alpha=1$. 
Assuming mergers in (active phase of) NSCs, where hierarchical mergers are probably most frequent, 
we set $m_\mathrm{max}=20\,\Msun$ as NSCs are mainly metal rich \citep[e.g.][]{Do18,Schodel20}, 
$v_\mathrm{esc}=1000\,\mathrm{km/s}$ typically expected for merging sites of binaries \citep{Tagawa19}, 
$\gamma_\mathrm{t}=2$ and $\gamma_\mathrm{q}=2$ as high- and equal-mass BHs are easier to merge in dynamical environments, 
and $\omega=0.1$ and $N_\mathrm{s}=4$ to reproduce frequent hierarchical mergers (Table~\ref{table:fraction_highg}, \citealt{Tagawa20_massgap}).

\subsection{Reconstruction of the spin distribution}

\label{sec:method_reconstruction_spin}

Here, we present a way to detect features for hierarchical mergers that possibly appear in the distribution of spins and masses. 

\subsubsection{Model characterizing the spin distribution}

\label{sec:model_xeff}

Given the universal trends of hierarchical mergers in the averaged spin magnitude as a function of masses for merging binaries 
($\S\,\ref{sec:result_pop_parameters}$), 
we investigate how well 
such trends can be reconstructed using a finite number of events. 
To do this, we replace the procedure above with a simple parametric analytic toy model, directly describing the distribution of the three variables 
(${\bm \theta}=\{\chi_\mathrm{eff},\chi_\mathrm{p},{m}_\mathrm{chirp}\}$) 
in terms of a set of the parameters (${\bm \lambda}$) as 
\begin{align}
\label{eq:p_theta_lambda}
p({\bm \theta}
|{\bm \lambda}) 
= N(\chi_\mathrm{typ}|\mu_\chi, \sigma_\chi) T[-1,1], 
\end{align}
where $N(x_0|x_1,x_2)$ represents the probability to return $x_0$ for the normal distribution with the mean $x_1$ and the standard deviation $x_2$, $T[-1,1]$ means to truncate the normal distribution to the range $[-1,1]$ and normalize $N$ so that the integral of $N$ in this range is $1$,  $\chi_\mathrm{typ}\equiv(\chi_\mathrm{p}^2 + \chi_\mathrm{eff}^2)^{1/2}$,  
\begin{align}
\label{eq:sigma_xeff1}
\mu_\chi=
\left\{
\begin{array}{l}
a_\mu m_\mathrm{chirp} + b_\mu  
~~~~~~\mathrm{for}~m_\mathrm{chirp}<m_{{\rm crit}}\\
b_\mu~~~~~~~~~~~~~~~~~~~~~~~\mathrm{for}~m_\mathrm{chirp}\geq m_{{\rm crit}}, 
\end{array}
\right.
\end{align}
and 
\begin{align}
\label{eq:sigma_xeff2}
\sigma_\chi=
\left\{
\begin{array}{l}
a_\sigma m_\mathrm{chirp} + b_\sigma  
~~~~~~\mathrm{for}~m_\mathrm{chirp}<m_{{\rm crit}}\\
b_\sigma~~~~~~~~~~~~~~~~~~~~~~~\mathrm{for}~m_\mathrm{chirp}\geq m_{{\rm crit}}. 
\end{array}
\right.
\end{align}
We use $\chi_\mathrm{typ}$ since it roughly represents the spin magnitudes of BHs in a binary. 
Hence, this model 
has five parameters ${\bm \lambda}=\{a_\mu, b_\mu, a_\sigma, b_\sigma, m_{{\rm crit}}\}$ characterizing the $\chi_\mathrm{typ}$ profile as a function of $m_\mathrm{chirp}$. 
The functional form of the model (eq.~\ref{eq:p_theta_lambda}) is motivated by the prediction that hierarchical mergers favor a plateau in the distribution of 
${\bar \chi}_\mathrm{typ}$ {\it vs.} $m_\mathrm{chirp}$
at high $m_\mathrm{chirp}$ as the BH spin magnitudes roughly converge to a constant value of $\sim 0.7$  
as a result of mergers with isotropic spin directions, 
while ${\bar \chi}_\mathrm{typ}$
roughly linearly approaches the value at the plateau from lower $m_\mathrm{chirp}$ 
according to Figs.~\ref{fig:x_mc_profile_dep} and \ref{fig_app:x_mc_profile_dep}.

We simply adopt the same functional form for $\sigma_\chi$ with $\nu_\chi$. 
Since the BHs formed from mergers typically have spins dominated by the orbital angular momentum of their progenitor binary (i.e. $\sim 0.7$), 
the dispersion in the $\chi_{\rm typ}$ distribution is expected to converge to a constant beyond $m_{\rm crit}$, producing a plateau. 
This motivates the functional form of Eq.~\eqref{eq:sigma_xeff2} to describe the relation between the spins and mass for hierarchical mergers.

The model parameters, ${\bm \lambda}$, are estimated from GW data through a Bayesian analysis, whose details are described in the next section.

\begin{table*}
\begin{center}
\caption{Fiducial values of our model parameters.}
\label{table:parameter_fiducial}
\hspace{-5mm}
\begin{tabular}{c|c}
\hline 
Parameter & Fiducial value \\
\hline\hline
The number of observed events & $N_\mathrm{obs}=1000$\\\hline
Frequency of mergers for high-mass binaries & $\gamma_\mathrm{t}=2$\\\hline
Frequency of mergers for equal-mass binaries & $\gamma_\mathrm{q}=2$\\\hline
The spin magnitudes for 1g BHs& $a_\mathrm{ave}=0$, $a_\mathrm{uni}=0$\\\hline
Maximum and minimum masses for 1g BHs& $m_\mathrm{max}=20\,\Msun$, $m_\mathrm{min}=5\,\Msun$\\\hline
Power-law exponent in the mass function for 1g BHs& $\alpha=1$\\\hline
Fraction of BHs that merges at each step & $\omega=0.1$\\\hline
Number of merger steps & $N_\mathrm{s}=4$\\\hline
Escape velocity of systems hosting BHs & $v_\mathrm{esc}=1000\,\mathrm{km/s}$\\\hline
The parameter for correlation between the steps and the redshift & $w_z=\infty$ (no correlation)\\\hline
\end{tabular}
\end{center}
\end{table*}

\subsubsection{Bayesian analysis}

\label{sec:bayesian_analysis}

To derive the posterior distribution of ${\bm \lambda}$ from a dataset $\{{\bm d}_i\}$,
$p( {\bm \lambda}|\{{\bm d}_i\})$, we use the Bayesian formalism as follows. 
Here, ${\bm d}_i$ encodes the measurable parameters ($\bm \theta$) and also includes their random noise in the $i^\mathrm{th}$ event. 
Bayes' rule gives 
\begin{align}
\label{eq:p_lambda_datai}
p( {\bm \lambda}|\{{\bm d}_i\}) = \frac{p( \{{\bm d}_i\}|{\bm \lambda})\pi ({\bm \lambda})}{p(\{{\bm d}_i\})},
\end{align}
where $p(\{ {\bm d}_i\}|{\bm \lambda})$ is the likelihood to obtain $\{{\bm d}_i\}$ for ${\bm \lambda}$, 
$\pi({\bm \lambda})$ is the prior probability for the model parameters ${\bm \lambda}$, and the evidence $p(\{{\bm d}_i\})$ is the integral of the numerator over all ${\bm \lambda}$.

We assume that each GW detection is independent, so that 
\begin{align}
\label{eq:p_lambda_alldata}
p( \{{\bm d}_i \}|{\bm \lambda})= \prod_{i=1}^{N_\mathrm{obs}} p({\bm d}_i|{\bm \lambda}). 
\end{align}

The probability of making observation $i$ is 
\begin{align}
\label{eq:p_datai}
p( {\bm d}_i|{\bm \lambda})=
\frac{\int d{\bm \theta} p({{\bm d}_i}|{\bm \theta})p({\bm \theta}|{\bm \lambda})}
{A({\bm \lambda})}, 
\end{align}
where 
the normalization factor $A({\bm \lambda})$ is given by 
\begin{align}
\label{eq:norm_a}
A({\bm \lambda})={\int_{{\bm d}>{\mathrm{threshold}}} d{\bm d} \int d {\bm \theta} p({\bm d}|{\bm \theta})p({\bm \theta}|{\bm \lambda})}\nonumber\\
={\int d{\bm \theta} p_\mathrm{det}({\bm \theta})p({\bm \theta}|{\bm \lambda})},
\end{align}
\begin{align}
\label{eq:p_detection}
p_\mathrm{det}({\bm \theta})=\int_{{\bm d}>{\mathrm{threshold}}}p({\bm d}|{\bm \theta})d{\bm d}
\end{align}
is the detection probability for a given set of parameters, 
and $``{\mathrm{threshold}}"$ denotes that the event ${\bm d}$ is detectable when ${\bm d}$ is above the threshold. 
To reduce computational costs, 
we assume that $A({\bm \lambda})$ is constant. 
This assumption does not affect our results as $A({\bm \lambda})$ varies by less than a factor of $1.1$ if the spin directions of BHs are assumed to be isotropic, meaning that the variation of $A({\bm \lambda})$ per each steps in the Monte Carlo method ($\S\,$\ref{sec:mcmc_method}) is negligible. 
This is because the detection probability is influenced only by $\chi_\mathrm{eff}$ by changing ${\bm \lambda}$ (see Eqs.~\ref{eq:rho0} and \ref{eq:p_theta_lambda}), and the reduction and enhancement of the detectable volume for mergers with negative and positive $\chi_\mathrm{eff}$ are mostly cancelled out.

The likelihood $p({\bm d}_i|{\bm \theta})$ 
can be rewritten in terms of the posterior probability density function (PDF) $p({\bm \theta}|{\bm d}_i)$ that is estimated in the analysis assuming prior $\pi({\bm \theta})$ as 
\begin{align}
\label{eq:p_detection2}
p({\bm d}_i|{\bm \theta})=\frac{p({\bm \theta}|{\bm d}_i)p({\bm d}_i)}{\pi({\bm \theta})}. 
\end{align}
The posterior PDF $p({\bm \theta}|{\bm d}_i)$ has information on errors, 
and it is often discretely sampled with $S_i$ samples from the posterior, $\{^j {\bm \theta}^{(i)}\}$, for $j\in[1,S_i]$. Because the samples are drawn according to the posterior, the parameter space volume associated with each sample is inversely proportional to the local PDF, $d^j \theta^{(i)} \propto [p(^j\theta ^{(i)}|d^{(i)})]^{-1}$, which allows us to replace the integral with a discrete sum \citep[e.g.][]{Mandel19,2020arXiv200705579V}. 
Overall, the posterior distribution of ${\bm \lambda}$ is given as 
\begin{align}
\label{eq:pf_lambda_di}
p&( {\bm \lambda}|\{{\bm d}_i \})\nonumber\\
&=\frac{\pi({\bm \lambda})}{p(\{{\bm d}_i\})}\prod_{i=1}^{{N_\mathrm{obs}}}
\frac{\frac{1}{S_i}\sum_{j=1}^{S_i}p(^j{\bm \theta}^{(i)}|{\bm \lambda})\frac{p({\bm d}_i)}{\pi({\bm \theta})}}{A({\bm \lambda})}\nonumber\\
&\propto 
\pi({\bm \lambda})\prod_{i=1}^{{N_\mathrm{obs}}}
\frac{\frac{1}{S_i}\sum_{j=1}^{S_i}p(^j{\bm \theta}^{(i)}|{\bm \lambda})\frac{1}{\pi({\bm \theta})}}{A({\bm \lambda})},
\end{align}
where we factor out the evidence factors $p(\{{\bm d}_i\})$ and $\prod_{i=1}^{{N_\mathrm{obs}}}p({\bm d}_i)$ since it is independent of ${\bm \lambda}$ and does not affect the relative values of the posterior $p( {\bm \lambda}|\{{\bm d}_i \})$. 
We use a flat prior distribution for $\pi ({\bm \lambda})$. 
We set $\pi ({\bm \theta})\propto d_L^2(z)$ 
following the standard priors used in the LIGO/Virgo analysis of individual events \citep{Veitch15}. 
We assume flat priors on $\chi_{\rm p}$ and $\chi_{\rm eff}$. Note that this is different from the LIGO/Virgo analysis which used uniform priors for the component spin magnitudes 
and they are appropriately transformed to priors for $\chi_{\rm p}$ and $\chi_{\rm eff}$. 
We set $S_i=3N_\mathrm{obs}$ so that we can take into account uncertainties whose probability is in the order of $\sim 1/N_\mathrm{obs}$.

\subsubsection{Markov chain Monte Carlo methods}

\label{sec:mcmc_method}

We calculate the posterior distribution (Eq.~\ref{eq:pf_lambda_di}) using Markov chain Monte Carlo (MCMC) methods. 
We track one chain for $10^7$ steps, 
set the first half to a burn-in period, check convergence by verifying that values for parameters after the burn-in period are oscillating around a constant average and dispersion. 
We adopt Metropolis-Hastings algorithm \citep[e.g.][]{Hastings1970}, and set a proposal distribution to the normal distribution with the values at each step as 
the means and the standard deviations for $a_\mu$, $b_\mu$, $a_\sigma$, $b_\sigma$, and $m_{{\rm crit}}$ to be 0.0001 $\Msun^{-1}$, 0.01, 0.0001 $\Msun^{-1}$, 0.01, and 1.0 $\Msun$, respectively. 
The standard deviations of the proposal distribution are roughly given by the typical standard deviations of the posterior distribution divided by $\sim4$ as this setting works well for convergence. 
We do not pose thinning to a posterior distribution as the autocorrelation for each variable between adjacent steps is already as small as $\lesssim 10^{-5}$. 
We restrict $m_{{\rm crit}}$ in the ranges from $m_\mathrm{min}$ to the maximum $m_\mathrm{chirp}$ among observed events.

\begin{figure*}
\begin{center}
\includegraphics[width=180mm]{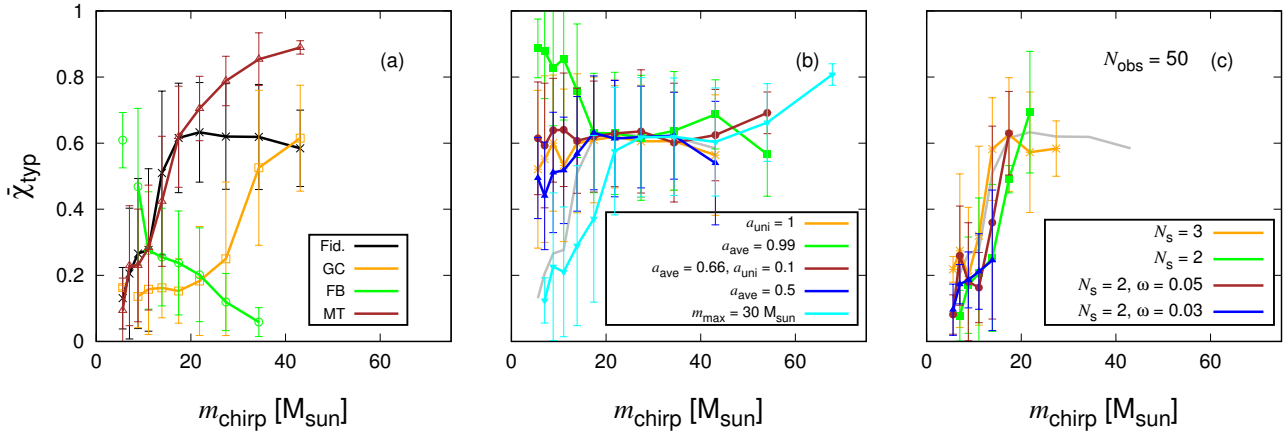}
\caption{
The mean dimensionless total spin ${\bar \chi}_\mathrm{typ}$ as a function of $m_\mathrm{chirp}$ for models~M1--M13 (Table~\ref{table:fraction_highg}). 
We use $N_\mathrm{obs}=10^3$ detectable mergers. 
In panels~(b)--(c), the profiles for model~M1 are presented by gray lines. 
Bars correspond to $1\sigma$ credible intervals. 
}
\label{fig:x_mc_profile_dep}
\end{center}
\end{figure*}

\begin{table*}
\begin{center}
\caption{
Properties of hierarchical mergers in our models. 
The first and second columns indicate the model number and its variation from the fiducial model (Table~\ref{table:parameter_fiducial}). 
The third and fourth columns show the fraction of high-g mergers among all and detectable mergers, respectively. The fifth column shows the maximum chirp mass ($m_\mathrm{chirp,max}$) among $N_\mathrm{obs}=10^3$ detectable mergers. 
The sixth and seventh columns show the average and the standard deviation of $\chi_\mathrm{p}$ among all merging pairs. 
}
\label{table:fraction_highg}
\hspace{-5mm}
\begin{tabular}{c|c|c|c|c|c|c}
\hline 
model&Parameter & high-g fraction & high-g detection fraction 
& $m_\mathrm{chirp,max} [\Msun]$ & ${\bar \chi}_\mathrm{p}$&$\sigma(\chi_\mathrm{p})$
\\\hline\hline
M1&Fiducial & 0.33 &0.68&56&0.17 & 0.26\\\hline
M2&Globular cluster (GC) & 0.063 & 0.17 & 44&0.030&0.13\\\hline
M3&Field binary (FB) & 0 & 0 & 33&0&0\\\hline
M4&Migration trap (MT) & 0.31 & 0.80 & 42&0&0\\\hline
M5&$a_\mathrm{uni}=1$ & 0.32 & 0.73 & 52&0.50&0.21\\\hline
M6&$a_\mathrm{ave}=0.99$ & 0.31 & 0.70 & 51&0.75&0.20\\\hline
M7&$a_\mathrm{ave}=0.66$, $a_\mathrm{uni}=0.1$ & 0.33 & 0.72 & 55&0.55&0.13\\\hline
M8&$a_\mathrm{ave}=0.5$ & 0.33 & 0.74 & 65&0.46&0.12\\\hline
M9&$m_\mathrm{max}=30\,\Msun$ & 0.35 & 0.73 & 70&0.18&0.26\\\hline
M10&$N_\mathrm{obs}=50$, $N_\mathrm{s}=3$& 0.25& 0.62 & 28&0.13&0.24\\\hline
M11&$N_\mathrm{obs}=50$, $N_\mathrm{s}=2$& 0.15& 0.28 & 24&0.077&0.19\\\hline
M12&$N_\mathrm{obs}=50$, $N_\mathrm{s}=2$, $\omega=0.05$& 0.077& 0.18 & 19&0.040&0.14\\\hline
M13&$N_\mathrm{obs}=50$, $N_\mathrm{s}=2$, $\omega=0.03$& 0.046& 0.14 & 19&0.023&0.11\\\hline
\end{tabular}
\end{center}
\end{table*}

\section{Results}

In \S\,\ref{sec:result_profiles_mock},
we investigate characteristic features in hierarchical mergers,
using our flexible tool (\S\,\ref{section:construction_gw_data}) to generate mock GW datasets for a large range of input parameter combinations.
In $\S\,\ref{sec:results_ligo_data}$, 
we analyze GW data observed in LIGO/Virgo O1--O3a. We first derive signatures and properties of hierarchical mergers ($\S\,\ref{sec:result_rec_x_obs}$), using the simple fitting formula for spin {\it vs.} chirp mass (\S\,\ref{sec:model_xeff}).  We then assess ($\S\,\ref{sec:results_bayesian_factor}$) how well the predictions in our mock GW catalogs and in our physical AGN disk models \citep{Tagawa20_massgap} in fact match these observed GW data.
Finally, in $\S\,$\ref{sec:result_rec_x}, 
we analyze mock GW data, and investigate how well the signatures of hierarchical models, described by the simple fitting formulae (${\bar \chi}_\mathrm{typ}$ {\it vs.} $m_\mathrm{chirp}$), can be recovered from future, larger GW catalogs.

\subsection{Profiles for average spin parameters}

\label{sec:result_profiles_mock}

\subsubsection{Dependence on population parameters}

\label{sec:result_pop_parameters}

We first show the parameter dependence of the ${\bar \chi}_\mathrm{typ}$ profile as a function of $m_\mathrm{chirp}$ using mock GW events, in which hierarchical mergers are assumed to be frequent. 
In Table~\ref{table:fraction_highg}, we list
the model varieties we have investigated.
These include the fiducial model (M1), and 12 different varieties (models~M2--M13). 
We examine different choices of 
the initial spin magnitudes (models~M5--M8) and the maximum mass of 1g BHs (model~M9), 
the fraction of hierarchical mergers (models~M10--M13), 
and the several parameter sets mimicking different populations (models~M2--M4, Table~\ref{table:parameters_populations}). 
We also investigate a variety of additional models in the appendix (models~M14--M28, Table~\ref{table_app:fraction_highg}).

Fig.~\ref{fig:x_mc_profile_dep} shows the profiles for models~M1--M13 (Table~\ref{table:fraction_highg}). 
For models in which hierarchical mergers are frequent (panels~(b) and (c) of Fig.~\ref{fig:x_mc_profile_dep} and Fig.~\ref{fig_app:x_mc_profile_dep}), 
there are 
universal trends for hierarchical mergers 
in the ${\bar \chi}_\mathrm{typ}$ profiles: 
(i) increase (or decrease) of ${\bar \chi}_\mathrm{typ}$ to $\sim 0.6$ at low $m_\mathrm{chirp}$. 
(ii) plateau of ${\bar \chi}_\mathrm{typ}$ with $\sim 0.6$ at high $m_\mathrm{chirp}$. 
Thus, the profile is roughly characterized by two lines if hierarchical mergers are frequent, mergers originate mostly from one population, and the typical spin magnitude for 1g BHs does not depend on their masses. 
The profile of ${\bar \chi}_\mathrm{typ}$ strongly depends on $a_\mathrm{ave}$, $a_\mathrm{uni}$, and $m_\mathrm{max}$ (Fig.~\ref{fig:x_mc_profile_dep}~b), while it is less affected by the other parameters (see Fig.~\ref{fig_app:x_mc_profile_dep}).

The typical value of ${\bar \chi}_\mathrm{typ}\sim 0.6$ at the plateau can be understood as follows. 
When masses and spin magnitudes between the primary and secondary BHs are similar ($m_\mathrm{1}\sim m_\mathrm{2}$ and $a_\mathrm{1}\sim a_\mathrm{2}\sim a_0$) 
and the directions of BH spins are isotropic, 
the typical magnitude of mass-weighted BH spins is 
\begin{equation}
|{\bm a}_\mathrm{w}|= 
\left\langle \left|\frac{m_\mathrm{1} {\bm a}_\mathrm{1} +m_\mathrm{2} {\bm a}_\mathrm{2}}{m_\mathrm{1}+m_\mathrm{2}}\right|\right\rangle
\sim\frac{\sqrt{7}}{3}a_0, 
\end{equation}
where $\langle\dots\rangle$ represents an average over the number of samples. 
If we approximate 
\begin{align}
{\bar \chi}_\mathrm{p} &\simeq 
\left\langle|{\bm a_0}| |\mathrm{cos}\theta| \right\rangle
\sim \frac{\pi}{4} a_0,\nonumber\\
\overline{|\chi_\mathrm{eff}|}&\simeq 
\left\langle|{\bm a_\mathrm{w}}| |\mathrm{sin}\theta|\right\rangle
\sim \frac{1}{2} |{\bm a_\mathrm{w}}|,
\end{align}
then 
\begin{align}
\label{eq:xtyp_xeff_xp}
{\bar \chi}_\mathrm{typ}&=\left ( \overline{|\chi_\mathrm{eff}|}^2+{\bar \chi}_\mathrm{p}^2\right)^{1/2}\nonumber\\
&\sim \left[\left(\frac{\sqrt{7}}{3}\frac{1}{2}\right)^2+\left(\frac{\pi}{4}\right)^2\right]^{1/2}a_0
\sim 0.90 a_0. 
\end{align}
Since merger remnants typically have spin magnitudes of $a_0\sim 0.7$ \citep{Buonanno08}, 
${\bar \chi}_\mathrm{typ}\sim 0.6$ for mergers among high-g BHs, which is roughly consistent with the value at the plateau (Figs.~\ref{fig:x_mc_profile_dep} and \ref{fig_app:x_mc_profile_dep}). 
Note that when $q\ll 1$, $|{\bm a}_\mathrm{w}|\sim a_0$ and so the average value is slightly enhanced to ${\bar \chi}_\mathrm{typ}\sim 0.93 a_0$.

\begin{figure*}
\begin{center}
\includegraphics[width=160mm]{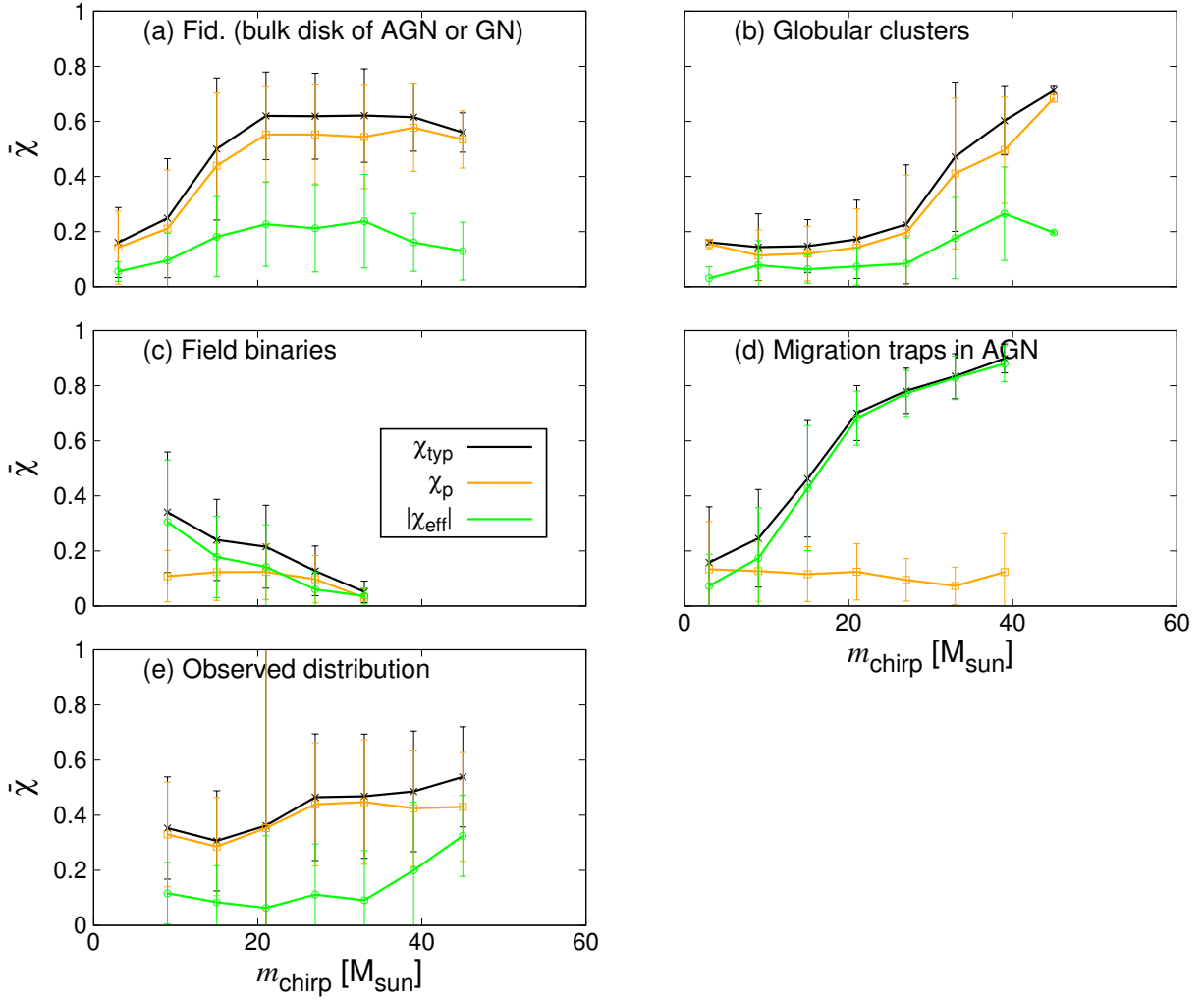}
\caption{
The profiles for the average spin parameters as a function of $m_\mathrm{chirp}$ for $10^3$ detectable mergers. 
Black, orange, and green lines represent the average of $\chi_\mathrm{typ}$, $\chi_\mathrm{p}$, and $|\chi_\mathrm{eff}|$, respectively. 
Panels~(a)--(e), respectively, present the distributions for mergers in AGN disks (M1), GCs (M2), FBs (M3), MTs (M4), and those observed by LIGO/Virgo O1--O3a. The averages for observed distributions (e) are calculated by averaging the medians of the parameters estimated in observed events. 
Bars correspond to $1\sigma$ credible intervals. 
}
\label{fig:x_mc_profile_pop}
\end{center}
\end{figure*}

As $m_\mathrm{max}$ increases, the bending point between the two lines increases (gray and cyan lines in Fig.~\ref{fig:x_mc_profile_dep}~b). This is because $m_\mathrm{max}$ determines the critical mass above which all merging BHs are of high generations with high spins of $\sim 0.7$. As the bending point is not influenced by the other parameters, the maximum mass of 1g BHs can be estimated from the bending point of the ${\bar \chi}_\mathrm{typ}$ profile. 
Note that since the bending points of the $\chi_\mathrm{p}$ and $\chi_\mathrm{eff}$ profiles are similar in shape to that of the $\chi_\mathrm{typ}$ profile for mergers with isotropic BH spins (Fig.~\ref{fig:x_mc_profile_pop}~a), either $\chi_\mathrm{typ}$, $\chi_\mathrm{p}$ or $\chi_\mathrm{eff}$ can constrain the maximum mass of 1g BHs if the profiles are reconstructed well.

Additionally, $a_\mathrm{ave}$ and $a_\mathrm{uni}$ influence ${\bar \chi}_\mathrm{typ}$ at the smallest values of $m_\mathrm{chirp}$ (Fig.~\ref{fig:x_mc_profile_dep}~b). 
This suggests that typical spin magnitudes of 1g BHs can be presumed by spins at small $m_\mathrm{chirp}$. 
However, note that ${\bar \chi}_\mathrm{typ}$ at small $m_\mathrm{chirp}$ is also influenced by the observational errors on $\chi_\mathrm{p}$ and $\chi_\mathrm{eff}$. 
Due to the smaller errors on $|\chi_\mathrm{eff}|$ compared to $\chi_\mathrm{p}$, $\overline{|\chi_\mathrm{eff}|}$ may constrain the typical spin values of 1g BHs more precisely using a number of events (green and orange lines in Fig.~\ref{fig:x_mc_profile_pop}~a). Note that ${\bar \chi}_\mathrm{p}>\overline{|\chi_\mathrm{eff}|}$ when the BH spins are isotropic due to their definition.  
In model~M7, the average and the dispersion of the spin magnitude for 1g BHs are set to be roughly the same as for the merger remnants. 
In such cases, the signatures of hierarchical mergers cannot be identified from the spin distributions (brown line in Fig.~\ref{fig:x_mc_profile_dep}~b). 
Also, for models in which the typical spin magnitude for 1g BHs are close to $\sim 0.7$ (e.g. models~M5 and M8), a large number of events are needed to detect the hierarchical merger signatures.

In Fig.~\ref{fig:x_mc_profile_dep}~(c), we can see how the features for hierarchical mergers in the $\chi_\mathrm{typ}$ profile are influenced by the fraction of hierarchical mergers for $N_\mathrm{obs}=50$. 
The plateau at high $m_\mathrm{chirp}$ is seen for $N_\mathrm{s}= 3$ (orange), while the rise of $\chi_\mathrm{typ}$ to $\sim 0.6$ at low $m_\mathrm{chirp}$ is seen for $N_\mathrm{s}=2$ with $\omega\geq 0.05$ (green and brown). 
These suggest that with $N_\mathrm{obs}=50$ 
the plateau and the rise of $\chi_\mathrm{typ}$ to $\sim0.6$ can be confirmed when the detection fraction of mergers of high-g BHs roughly exceeds $\sim 0.5$ and $\sim 0.15$, respectively (models~M10, M12, Table~\ref{table:fraction_highg}).

To summarize, the profile of ${\bar \chi}_\mathrm{typ}$ is mostly affected only by $a_\mathrm{ave}$, $a_\mathrm{uni}$, and $m_\mathrm{max}$, 
while the other parameters may affect the maximum $m_\mathrm{chirp}$ or the frequency of high-g mergers (Tables~\ref{table:fraction_highg} and \ref{table_app:fraction_highg}).

\begin{table}
\begin{center}
\caption{
Adopted parameter values for several populations. 
The differences with respect to the fiducial model (Table~\ref{table:parameter_fiducial}) are listed. 
}
\label{table:parameters_populations}
\hspace{-5mm}
\begin{tabular}{c|c}
\hline \hline
\multicolumn{2}{c}{Globular cluster (GC)}  \\\hline
1& $m_\mathrm{max}=45\,\Msun$ \\\hline
2& $N_\mathrm{s}=2$\\\hline
3& $\omega=0.03$\\\hline
4& $v_\mathrm{esc}=30\,\mathrm{km/s}$\\\hline\hline
\multicolumn{2}{c}{Field binary (FB)}  \\\hline
1& $m_\mathrm{max}=45\,\Msun$\\\hline
2& $N_\mathrm{s}=1$\\\hline
3& $\theta_1=\theta_2=0$\\\hline
4& $a_\mathrm{uni}$
follows Eq.~\eqref{eq:a_fb} 
\\\hline\hline
\multicolumn{2}{c}{Migration trap (MT)}  \\\hline
1& $\theta_1=\theta_2=0$\\\hline
\end{tabular}
\end{center}
\end{table}

\begin{figure*}
\begin{center}
\includegraphics[width=160mm]{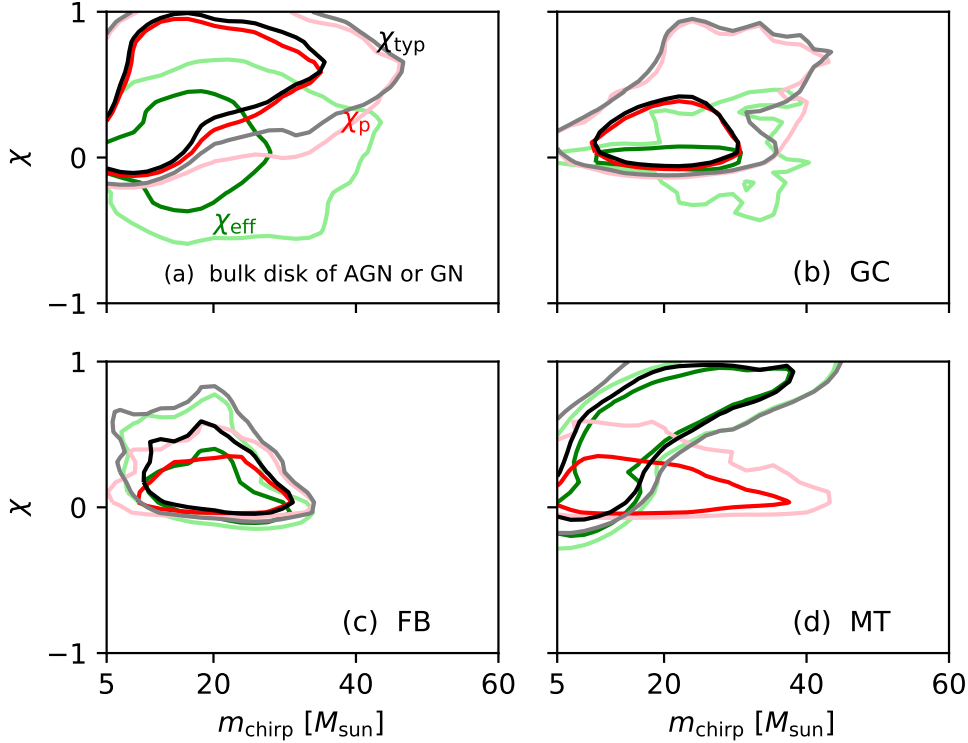}
\caption{
The 90 (dark lines) and 99 (light lines) percentile distributions in 
the spin {\it vs.} $m_\mathrm{chirp}$ plane. 
Black, red, and green lines represent the distributions of $\chi_\mathrm{typ}$, $\chi_\mathrm{p}$, and $\chi_\mathrm{eff}$, respectively. 
Panels~(a), (b), (c), and (d) show distributions for mergers in AGNs (bulk disks), GCs, FBs, and MTs, respectively. 
}
\label{fig:dist_mc_x_pop}
\end{center}
\end{figure*}

\begin{figure*}
\begin{center}
\includegraphics[width=160mm]{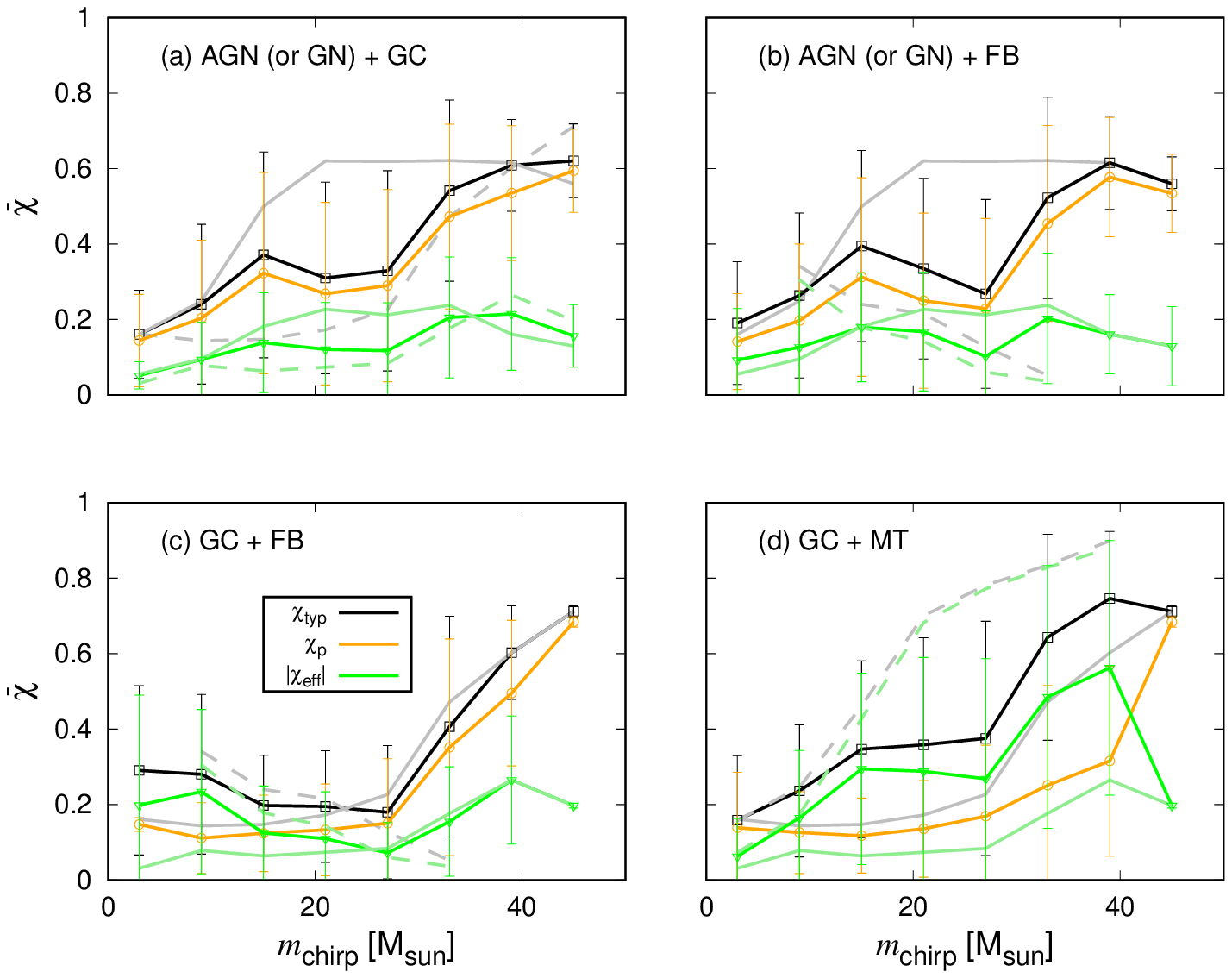}
\caption{
Same as Fig.~\ref{fig:x_mc_profile_pop}, but for mixture of two models. Both channels contribute to 1000 mergers, 
and contributing two models are specified in the upper left of each panel. "AGN", "GC", "FB", and "MT" represent mergers in the AGN disks, GCs, FBs, and MTs, respectively. 
Thin and dashed lines present the $\chi_\mathrm{eff}$ and $\chi_\mathrm{typ}$ profiles only for the former and the latter populations in the legend, respectively. 
}
\label{fig:x_mc_data_mix}
\end{center}
\end{figure*}

\subsubsection{Contribution from multiple populations}
\label{sec:results_several_pop}

In the fiducial model (M1), the parameter values (Table~\ref{table:parameter_fiducial}) are roughly adjusted to reproduce properties of mergers in AGN disks outside of MTs \citep{Tagawa20_massgap} or NSCs. 
The $\chi_\mathrm{typ}$ profile is similar 
but the $m_\mathrm{chirp}$ distribution is different between 
the fiducial model and physically motivated models derived in \citet{Tagawa20_massgap}. 
The former is because the profile is characterized by the few parameters ($m_\mathrm{max}$, $a_\mathrm{uni}$, $a_\mathrm{ave}$) as found in $\S\,\ref{sec:result_profiles_mock}$, while the latter is because the $m_\mathrm{chirp}$ distributions are affected by how BHs pair with other BHs and merge in AGN disks.

In this section, we additionally consider the spin distributions 
for mergers typically expected in several environments, including GCs, FBs, and MTs of AGN disks. 
Values of the parameters adopted to mimic these populations are listed in Table~\ref{table:parameters_populations}. 
Figs.~\ref{fig:dist_mc_x_pop} and \ref{fig:x_mc_profile_pop}, and panel~(a) in Fig.~\ref{fig:x_mc_profile_dep} present 
the distributions and the profiles of the spin parameters ($\chi_\mathrm{typ}$, $\chi_\mathrm{p}$, and $\chi_\mathrm{eff}$) as functions of $m_\mathrm{chirp}$ for these populations. 
Fig.~\ref{fig:x_mc_data_mix} is the same as Fig.~\ref{fig:x_mc_profile_pop}, but mergers are contributed by a mixture of two populations. 
Some contribution from multiple populations to the observed events is also favored by the analysis in \citet{Zevin2020_multiple}.

For mergers in GCs, 
we set lower escape velocity $v_\mathrm{esc}=30\,\mathrm{km/s}$, $N_\mathrm{s}=2$ and $\omega=0.03$ to reproduce the detection fraction of hierarchical mergers of $\sim 10$--$20\%$, which is predicted by theoretical studies  \citep[e.g.][Table~\ref{table:fraction_highg}]{OLeary16,Rodriguez19_hierarchical}. We chose higher $m_\mathrm{max}=45\,\Msun$ as GCs are composed of metal-poor stars \citep[e.g.][]{Peng06,Leaman13,Brodie14}; other parameters are the same as those for AGN disks. 
Note that $m_\mathrm{max}$ in metal-poor environments is uncertain due to uncertainties on the reaction rate of carbon burning \citep{Farmer19} and the enhancement of the helium core mass by rotational mixing \citep{Chatzopoulos12,Yoon12,Vink2021}.

Due to higher $m_\mathrm{max}$, 
${\bar \chi}_\mathrm{typ}$ continues to increase until higher $m_\mathrm{chirp}$ (panel~b in Fig.~\ref{fig:x_mc_profile_pop}, see also \citealt[][]{Rodriguez18PRL}) compared to the fiducial model (panel~a). 
Also, 90 percentile regions are distributed around $\chi_\mathrm{eff}\sim 0$ and $\chi_\mathrm{p}\sim 0$ (Fig.~\ref{fig:dist_mc_x_pop}~b) as a large fraction of mergers are among 1g BHs. 
Thus, the distribution of ${\bar \chi}_\mathrm{typ}$ at low $m_\mathrm{chirp}$ is clearly different between mergers in AGN disks and GCs, mainly due to the difference of $m_\mathrm{max}$ and the fraction of mergers among high-g BHs. 
If mergers are comparably contributed both by GCs and AGN disks, 
steep increase of ${\bar \chi}_\mathrm{typ}$ against $m_{\rm chirp}$ appears twice (panel~a in Fig.~\ref{fig:x_mc_data_mix}). 
Thus, mixture of these populations can be discriminated by analyzing the spin distribution. 
Note that the intermediate line between the two increases in the ${\bar \chi}_\mathrm{typ}$ profile is roughly characterized by the ratio of mergers from AGN disks and GCs. Hence, the contribution from multiple populations would be distinguishable by analyzing the profile by using a number of GW events.

For mergers among FBs, we set $N_\mathrm{s}=1$ and $m_\mathrm{max}=45\,\Msun$. 
Although BH spin distributions are highly uncertain, we refer to \citet{Bavera19} who proposed that ${\bar \chi}_\mathrm{eff}$ is high at low $m_\mathrm{chirp}$ of $\lesssim 10-20\,\Msun$ as low-mass progenitors have enough time to be tidally spun up. 
We assume that $a_\mathrm{uni}$ follows 
\begin{align}
\label{eq:a_fb}
a_\mathrm{uni}=
\left\{
\begin{array}{l}1~~\mathrm{for}~~~ m_\mathrm{1g}\leq 15\,\Msun \\
(30\,\Msun- m_\mathrm{1g})/15\,\Msun \\
~~~~~~~~~~~~\mathrm{for}~~~ 15\,\Msun \leq m_\mathrm{1g}\leq 30\,\Msun \\
0~~\mathrm{for}~~~ 30\,\Msun \leq m_\mathrm{1g} .
\end{array}
\right.
\end{align}
BH spins are assumed to be always aligned with the orbital angular momentum of binaries, although 
we do not always expect spins to be aligned \citep[e.g.][]{Kalogera2000,Rodriguez2016_kicks}. 
In such a setting, $\overline{|\chi_\mathrm{eff}|}$ decreases as $m_\mathrm{chirp}$ increases (panels~c of Figs.~\ref{fig:dist_mc_x_pop} and \ref{fig:x_mc_profile_pop}). 
Also, non-zero $\chi_\mathrm{p}$ is due to assumed observational errors (orange line in Fig.~\ref{fig:x_mc_profile_pop}~c). 
The profile expected for the binary evolution channel is significantly different from those expected for the other channels. 
If mergers arise comparably from FBs and GCs, $\overline{|\chi_\mathrm{eff}|}$ exceeds ${\bar \chi}_\mathrm{p}$ at low $m_\mathrm{chirp}$ (panel~c of Fig.~\ref{fig:x_mc_data_mix}). 
As contribution from mergers in FBs enhances $\overline{|\chi_\mathrm{eff}|}$ relative to ${\bar \chi}_\mathrm{p}$ at low $m_\mathrm{chirp}$, we could constrain the contribution from FBs using the ratio of $\overline{|\chi_\mathrm{eff}|}$ to ${\bar \chi}_\mathrm{p}$. 
Observed events so far suggest that $|\chi_\mathrm{eff}|$ is typically lower than $\chi_\mathrm{p}$ at low $m_\mathrm{chirp}$ (panel~e of Fig.~\ref{fig:x_mc_profile_pop}), implying that the contribution to the observed mergers from FBs is minor, unless adopted spins for 1g BHs need significant revisions.

For mergers in MTs, we assume that parameters are the same as in the fiducial model (Table\,\ref{table:parameter_fiducial}), while BH spins are always aligned with the orbital angular momentum of the binaries. 
Such alignment is expected for binaries in MTs where randomization of the binary orbital angular momentum directions by binary-single interactions is inefficient due to rapid hardening and merger caused by gas dynamical friction (unlike in gaps formed further out in the disk, where these interactions were found to be very important by \citealt{Tagawa20b_spin}), 
and so the BH spins are aligned with circumbinary disks due to the Bardeeen-Petterson effect \citep{Bardeen75}, and circumbinary disks are aligned with the binaries due to viscous torque \citep[e.g.][]{Moody19}. 
Here, we assume that the orbital angular momentum directions of binaries are the same as that of the AGN disk referring to \citet{Lubow99}, which is different from the assumption (anti-alignment with $50\%$) adopted in \citet{Yang19b_PRL}. 
In this model, the $\chi_\mathrm{p}$ and $|\chi_\mathrm{eff}|$ distributions are significantly different from those in the other models (panels~d of Figs.~\ref{fig:dist_mc_x_pop} and \ref{fig:x_mc_profile_pop}). 
The value of $\chi_\mathrm{eff}$ at high $m_\mathrm{chirp}$ is typically high, while $\chi_\mathrm{p}$ is low. 
When mergers originate comparably in MTs and GCs, $\overline{|\chi_\mathrm{eff}|}$ significantly exceeds ${\bar \chi}_\mathrm{p}$ in a wide range of $m_\mathrm{chirp}$ (Fig.~\ref{fig:x_mc_data_mix}~d). 
As $\overline{|\chi_\mathrm{eff}|}$ is typically lower than ${\bar \chi}_\mathrm{p}$ in the observed events in all $m_\mathrm{chirp}$ bins (Fig.~\ref{fig:x_mc_data_mix}~e), the contribution from MTs to the detected mergers is probably minor.

\begin{figure*}
\begin{center}
\includegraphics[width=150mm]{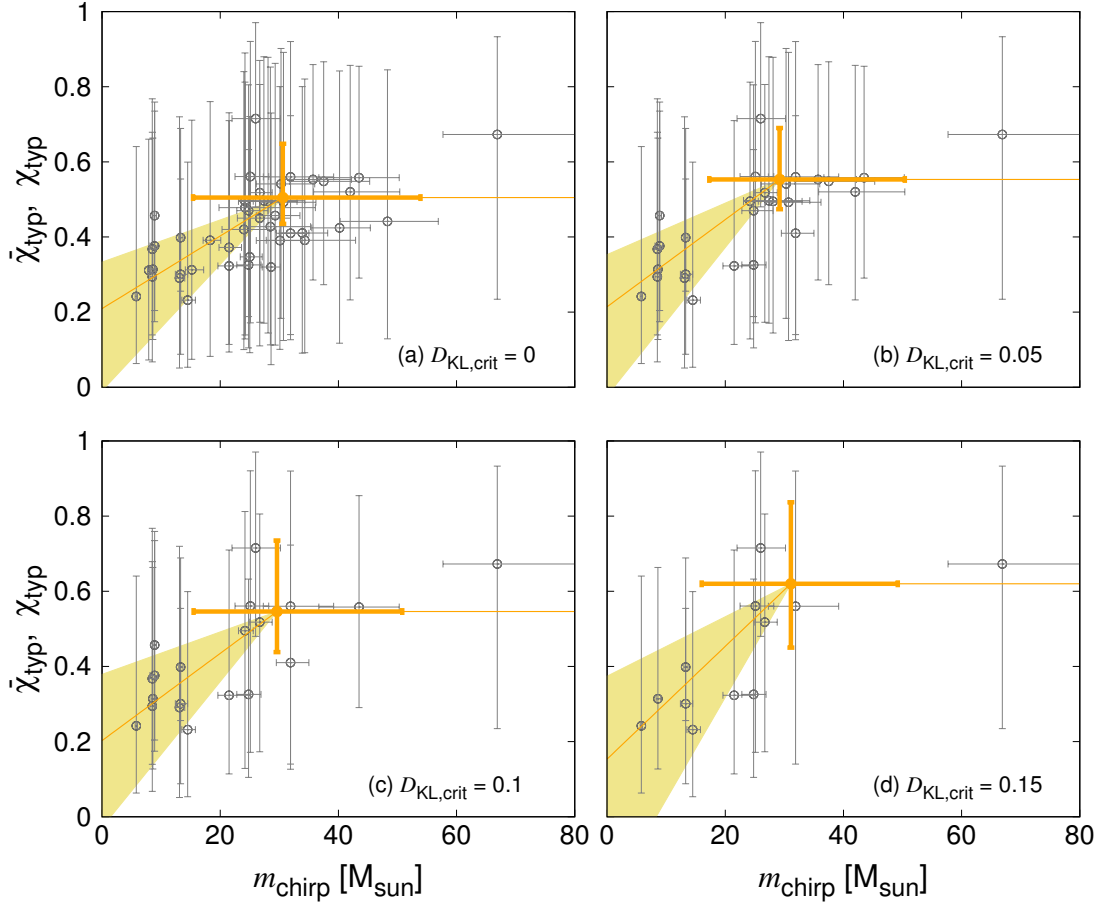}
\caption{
The ${\bar \chi}_\mathrm{typ}$ profile as a function of $m_\mathrm{chirp}$ constructed using the GW data observed in LIGO/Virgo O1--O3a. 
Orange line presents the recovered profile using the model described in $\S\,\ref{sec:method_reconstruction_spin}$ (Eq.~\ref{eq:sigma_xeff1}). 
Bars and shaded regions represent $1\sigma$ credible intervals 
for ${\bar \chi}_\mathrm{typ}$ at the plateau $b_\mu$, 
the critical chirp mass at the bending point of the profile $m_{\mathrm{crit}}$, 
and the slope of the ${\bar \chi}_\mathrm{typ}$ at a lower chirp mass $a_\mu$. 
Panels (a), (b), (c), and (d) presents results for 
events with $D_\mathrm{KL}\geq$ 0, 0.05, 0.1, and 0.15, respectively. 
Gray plots are the same as those in Fig.~\ref{fig_app:mc_x_obs}~(c). 
}
\label{fig:x_mc_construct_obs}
\end{center}
\end{figure*}

\subsection{Application to LIGO/Virgo O1--O3a data}

\label{sec:results_ligo_data}

\subsubsection{Reconstruction of spin profiles}

\label{sec:result_rec_x_obs}

We analyze the GW data observed in LIGO/Virgo O1--O3a reported by \citet{TheLIGO18} and \citet{LIGO20_O3_Catalog}. 
Although $\chi_\mathrm{p}$ and $\chi_\mathrm{typ}$ suffer large uncertainties 
(e.g. Fig.~\ref{fig_app:mc_x_obs}), 
their median values indicate a positive correlation with $m_\mathrm{chirp}$. 
Such positive correlation is, if confirmed, consistent with the growth of BH spin magnitudes by hierarchical mergers as presented in Figs.~\ref{fig:x_mc_profile_dep}, \ref{fig:x_mc_profile_pop}, and \ref{fig_app:x_mc_profile_dep}.

To confirm the features in the $\chi$-profiles due to hierarchical mergers, 
we reconstruct the ${\bar \chi}_\mathrm{typ}$ profile from the observed GW data 
in the way described in $\S\,\ref{sec:method_reconstruction_spin}$. 
We discretize the posteriors for $m_{\rm chirp}$, $\chi_{\rm eff}$, and $\chi_{\rm p}$ with 20, 40, and 20 bins in the ranges from the minimum to the maximum of posteriors for $m_{{\rm chirp},i}$, from -1 to 1, and from 0 to 1, respectively. 
Note that the prior and posterior distributions for some events 
are similar 
to each other, which means that $\chi_\mathrm{p}$ is less constrained by the waveforms. 
To exclude events in which $\chi_\mathrm{p}$ are not well estimated, 
we only use events in which the Kullback-Leibler (KL) divergence between prior and posterior samples evaluated using heuristic estimates of $\chi_\mathrm{p}$ ($D_\mathrm{KL}$) exceeds a critical value of $D_\mathrm{KL,cri}=0$, 0.05, 0.1, 0.15, or 0.2. 
We consider that $\chi_\mathrm{p}$ for events with non-zero $D_\mathrm{KL}$ is statistically useful to understand the spin distribution. 
We use the events with $m_2\geq 5\,\Msun$ 
provided in \citet{LIGO20_O12_HP} and \citet{LIGO21_O3_HP} 
as we do not model mergers of neutron stars. 
Then, the number of events with $D_\mathrm{KL}\geq 0$, 0.05, 0.1, 0.15, and 0.2 are 44, 28, 20, 12, and 7, respectively. 
We present $1\sigma$ errors on the estimated parameters below unless stated otherwise.

The reconstructed ${\bar \chi}_\mathrm{typ}$ profiles for 
$D_\mathrm{KL,cri}=0$, 0.05, 0.1, and 0.15 are, respectively, presented by orange lines in panels (a)--(d) of Fig.~\ref{fig:x_mc_construct_obs}, 
and the posterior distributions and correlations of the reconstructed parameters for $D_\mathrm{KL,cri}=0$ are presented in Fig.~\ref{fig_app:mc_dist} in the Appendix. 
For $D_\mathrm{KL,cri}=0$, 0.05, 0.1, 0.15, and 0.20, respectively,
${\bar \chi}_\mathrm{typ}$ at the plateau is 
$b_\mu=0.51^{+0.14}_{-0.07}$, $0.55^{+0.14}_{-0.08}$, $0.55^{+0.19}_{-0.11}$, $0.62^{+0.22}_{-0.23}$, and $0.66^{+0.20}_{-0.18}$, 
the critical chirp mass at the bending point of the ${\bar \chi}_\mathrm{typ}$ profile is $m_{\mathrm{crit}}=31^{+23}_{-15}\,\Msun$, $29^{+21}_{-12}\,\Msun$, $30^{+21}_{-14}\,\Msun$, $31^{+18}_{-15}\,\Msun$, and $36^{+17}_{-16}\,\Msun$, 
and the slope of ${\bar \chi}_\mathrm{typ}$ at $m_\mathrm{chirp}<m_{\mathrm{crit}}$ is 
$a_\mu=10 ^{+7}_{-4}\times 10^{-3}\,\Msun^{-1}$, 
$12 ^{+8}_{-5}\times 10^{-3}\,\Msun^{-1}$, 
$12 ^{+7}_{-6}\times 10^{-3}\,\Msun^{-1}$, 
$15 ^{+12}_{-7}\times 10^{-3}\,\Msun^{-1}$, and 
$15 ^{+8}_{-6}\times 10^{-3}\,\Msun^{-1}$. 
To understand the influence of GW190521, which seems to have a large impact on spin distributions due to its large mass and $\chi_{\rm p}$, 
we repeated our analysis excluding this event. 
In this case, for $D_\mathrm{KL,cri}=0$, 0.05, and 0.1, respectively,
$b_\mu=0.50^{+0.15}_{-0.07}$, $0.52^{+0.14}_{-0.08}$, and $0.59^{+0.24}_{-0.14}$, 
$m_{\mathrm{crit}}=36^{+19}_{-14}\,\Msun$, $25^{+23}_{-11}\,\Msun$, and $37^{+19}_{-17}\,\Msun$, 
and 
$a_\mu=8 ^{+4}_{-4}\times 10^{-3}\,\Msun^{-1}$, 
$12 ^{+10}_{-6}\times 10^{-3}\,\Msun^{-1}$, and 
$12 ^{+4}_{-5}\times 10^{-3}\,\Msun^{-1}$, 
while for $D_\mathrm{KL,cri}=0.15$ and $0.2$, the parameters are not well determined due to the small number of events. For $D_{\rm LK,crit}\leq0.1$, the evaluated values of the parameters are similar with and without GW190521. 

The positive value of the slope ($a_\mu$), i.e., the increase of ${\bar \chi}_\mathrm{typ}$ at low $m_\mathrm{chirp}$ is confirmed with $\gtrsim 2\sigma$ confidence, 
which is a tell-tale sign of frequent hierarchical mergers. 
Also, according to the analysis in $\S\,\ref{sec:result_profiles_mock}$, 
the detection of the rise of ${\bar \chi}_\mathrm{typ}$ at low $m_\mathrm{chirp}$ with $N_\mathrm{obs}=50$ roughly requires that the detection fraction of mergers of high-g BHs exceeds $\sim 0.15$. As the number of events is smaller than 50 (e.g. $N_\mathrm{obs}=28$ for $D_\mathrm{KL,cri}=0.05$), 
the high-g detection fraction would be even higher than $\sim 0.15$. 
Thus, hierarchical mergers are preferred from the analysis. 
Note that accretion can also produce a positive correlation, but $|\chi_\mathrm{eff}|>\chi_\mathrm{p}$ is predicted in such cases, similarly to mergers in MTs (panel~d of Fig.~\ref{fig:x_mc_profile_pop}). As $|\chi_\mathrm{eff}|<\chi_\mathrm{p}$ is predicted by GW observations (panel~e of Fig.~\ref{fig:x_mc_profile_pop}), accretion is disfavored as a process enhancing the BH spin magnitudes.

For $D_\mathrm{KL,cri}=$ 0.05, 0.1, 0.15, and 0.20 (panels~b, c, and d of Fig.~\ref{fig:x_mc_construct_obs}), the value of ${\bar \chi}_\mathrm{typ}$ at the plateau ($b_\mu \sim 0.6$) is consistent with that expected from hierarchical mergers ($\sim 0.6$), which possibly supports frequent hierarchical mergers with the high-g detection fraction to be $\gtrsim 0.5$ ($\S\,$\ref{sec:result_pop_parameters}). 
On the other hand, for $D_\mathrm{KL,cri}=0$, $b_\mu \sim 0.5$, which is somewhat lower than the expected value of $0.6$. 
This is presumably because $\chi_\mathrm{p}$ values for 
events with $D_\mathrm{KL}\leq 0.05$ are not well constrained and just reflect assumed priors. 
Also, note that events with high $\chi_\mathrm{p}$ might tend to be missed as the waveform for large $\chi_\mathrm{p}$ \citep{Apostolatos94,Kidder95,Pratten20} or spin \citep{Kesden10,Gerosa19} 
mergers often accompany strong amplitude modulation, reducing SNRs.

Here, ${\bar \chi}_\mathrm{typ}$ at $m_\mathrm{chirp}=m_\mathrm{min}$ is closely related to the typical spin magnitude for 1g BHs (Fig.~\ref{fig:x_mc_profile_dep}~b). 
If we assume the median values 
for ${\bar \chi}_\mathrm{typ}$ and $m_{\mathrm{crit}}$, 
${\bar \chi}_\mathrm{typ}$ at $m_\mathrm{chirp}= 5\,\Msun$ is 
$0.26^{+0.10}_{-0.18}$, 
$0.27^{+0.12}_{-0.20}$, 
$0.26^{+0.14}_{-0.19}$, 
$0.23^{+0.18}_{-0.32}$, and 
$0.29^{+0.20}_{-0.25}$ 
for $D_\mathrm{KL,cri}=0$, 0.05, 0.1, 0.15, and 0.20, respectively. 
These suggest that 1g BHs typically have $a \lesssim 0.4$. 
Since this value is effectively enhanced by the observational errors on ${\bar \chi}_\mathrm{p}$, 
the estimated typical spin magnitude of 1g BHs is still consistent with $\sim 0$ as predicted by stellar evolution models \citep{Fuller19_massive}, which is also verified later ($\S\,$\ref{sec:result_rec_x}).

The critical chirp mass at the bending point of the ${\bar \chi}_\mathrm{typ}$ profile ($m_{\mathrm{crit}}$) is related to the maximum mass of 1g BHs (Fig.~\ref{fig:x_mc_profile_dep}~f). 
The analysis loosely constrains the parameter to $m_{\mathrm{crit}}\sim 15$--$50\,\Msun$, from which we discuss in $\S\,\ref{sec:result_rec_x}$ that the maximum mass of 1g BHs is estimated to be $\sim 20$--$60\,\Msun$. 
However, it needs a caution that $m_{\mathrm{crit}}$ is restricted from $5\,\Msun$ to the maximum chirp mass among the event ($\sim 67\,\Msun$) in this analysis, which may artificially produce the bending point and the plateau. 
To confidently confirm the plateau, $m_{\mathrm{crit}}$ needs to be precisely constrained compared to the allowed range for $m_{\mathrm{crit}}$ of $5$--$67\,\Msun$, 
which would require further events (see also $\S\,\ref{sec:result_rec_x}$).

\begin{table*}
\begin{center}
\caption{
The parameters of model~A which are different from each population model (shown in Table~\ref{table:parameters_populations}) and the logarithm of their Bayes factor $K_\mathrm{A,B}$ relative to the fiducial model ("B"). 
The Bayes factors for the three parameters with $D_\mathrm{KL,cri}=0$, 0.05, 0.1, 0.15, and 0.2, and that for the two parameters are presented from the second to seventh columns. 
We highlight the models with positive Bayes factors in the five rightmost columns in boldface. 
}
\label{table:bayesian_values}
\hspace{-5mm}
\begin{tabular}{c|c|c|c|c|c|c}
\hline
Parameters&\multicolumn{5}{c}{$\mathrm{log}_{10} K_\mathrm{A,B}$ in 3D} 
&$\mathrm{log}_{10} K_\mathrm{A,B}$ in 2D\\\hline
&$D_\mathrm{KL,cri}=0$&$D_\mathrm{KL,cri}=0.05$&$D_\mathrm{KL,cri}=0.1$&$D_\mathrm{KL,cri}=0.15$&$D_\mathrm{KL,cri}=0.2$&-\\\hline \hline
\multicolumn{7}{c}{AGN disk or NSC}  \\\hline
$m_\mathrm{max}=15\,\Msun$& -6.9&-4.0&-1.8&-3.0&-1.5&-7.2\\\hline
$m_\mathrm{max}=25\,\Msun$& 1.3&-0.05&-1.1&-0.31&-0.27&{\bf 1.3}\\\hline
$m_\mathrm{max}=30\,\Msun$& -0.9&-2.0&-2.7&-1.4&-0.90&{\bf 0.085}\\\hline
$m_\mathrm{max}=35\,\Msun$& -3.2&-3.7&-4.6&-2.4&-1.6&-1.8\\\hline
$m_\mathrm{max}=45\,\Msun$& -8.2&-7.6&-7.4&-4.2&-2.6&-5.7\\\hline
$\alpha=2$& -0.065&{\bf 0.35}&{\bf 0.65}&{\bf 0.065}&{\bf 0.29}&-0.56\\\hline
$\alpha=2$, $m_\mathrm{max}=15\,\Msun$& -12&-8.3&-4.5&-4.1&-3.3&-15\\\hline
$\alpha=2$, $m_\mathrm{max}=25\,\Msun$& 1.5&{\bf 0.83}&{\bf 0.14}&-0.060&{\bf 0.13}&{\bf 1.4}\\\hline
$\alpha=2$, $m_\mathrm{max}=30\,\Msun$& 1.4&{\bf 0.12}&-1.0&-0.58&-0.33&{\bf 1.9}\\\hline
$\alpha=2$, $m_\mathrm{max}=35\,\Msun$&0.11&-1.1&-2.0&-1.2&-0.57 &{\bf 0.72}\\\hline
$\alpha=2$, $m_\mathrm{max}=45\,\Msun$& -2.7&-3.1&-3.4&-2.0&-0.87&-1.3\\\hline
$a_\mathrm{uni}=1$&-2.2&-1.7&-2.1&-1.7&-1.0&-1.3\\\hline
$a_\mathrm{ave}=0.3$& 0.51& {\bf 0.53}&{\bf 0.042}&-0.21&{\bf 0.062}&-0.035\\\hline
$a_\mathrm{ave}=0.5$&-1.5&-0.85&-1.2&-1.2&-0.67&-1.1 \\\hline
$a_\mathrm{ave}=0.7$&-5.5&-4.1&-4.3&-3.0&-2.0&-2.4 \\\hline
$\sigma_\mathrm{p}=0.3$& -0.021&{\bf 0.053}&-0.059&-0.15&-0.16&-0.24 \\\hline
$\sigma_\mathrm{p}=0.4$& 0.37&{\bf 0.42}&-0.075&-0.20&-0.072&{\bf 0.40} \\\hline
$v_\mathrm{esc}=30\,\mathrm{km/s}$& -14&-8.7&-6.0&-5.7&-5.7&-13\\\hline
$w=0.03$& -26&-23&-12&-13&-8.3&-30\\\hline
$w=0.2$& -3.2&-2.4&-3.1&-1.8&-0.81&-2.1\\\hline
$\gamma_\mathrm{t}=0$& -21&-16&-11&-11&-8.2&-25\\\hline
$\gamma_\mathrm{t}=4$& -3.7&-2.8&-3.1&-1.7&-0.64&-3.2\\\hline
$\gamma_\mathrm{q}=0$& -6.5&-4.5&-2.3&-2.8&-2.0&-6.7\\\hline
$\gamma_\mathrm{q}=4$& 0.39&{\bf 0.12}&-0.26&-0.18&-0.058&{\bf 0.64}\\\hline
$N_\mathrm{s}=2$& -49&-31&-25&-29&-25&-48\\\hline
$N_\mathrm{s}=2$, $m_\mathrm{max}=30\,\Msun$& -0.55&-3.3&-5.7&-2.9&-4.0&-0.84\\\hline
$N_\mathrm{s}=2$, $m_\mathrm{max}=45\,\Msun$&-3.6&-6.1&-7.9&-3.9&-3.6 &-0.95\\\hline
$N_\mathrm{s}=2$, $m_\mathrm{max}=60\,\Msun$& -14&-13&-12&-6.9&-5.0&-9.5\\\hline
$N_\mathrm{s}=3$&-4.1&-3.0&-1.5&-1.8&-2.0&-5.9 \\\hline
$N_\mathrm{s}=3$, $m_\mathrm{max}=25\,\Msun$& 1.0&-0.55&-1.2&-0.56&-0.77&{\bf 0.81}\\\hline
$N_\mathrm{s}=3$, $m_\mathrm{max}=30\,\Msun$&1.9&-0.11&-1.7&-0.61&-0.77&{\bf 2.1}\\\hline
$N_\mathrm{s}=3$, $m_\mathrm{max}=35\,\Msun$&0.24&-2.1&-3.8&-1.8&-1.5&{\bf 1.3} \\\hline
$N_\mathrm{s}=3$, $m_\mathrm{max}=45\,\Msun$& -6.7&-7.2&-7.7&-4.2&-2.9&-3.7\\\hline
$N_\mathrm{s}=5$& 0.084&{\bf 0.16}&-0.50&-0.48&{\bf 0.14}& {\bf 0.44} \\\hline
$N_\mathrm{s}=5$, $m_\mathrm{max}=15\,\Msun$&-0.96&-0.17&-0.091&-0.76&-0.065& -0.68\\\hline
$N_\mathrm{s}=5$, $m_\mathrm{max}=30\,\Msun$&-3.6&-3.4&-3.8&-2.1&-1.1& -2.6\\\hline
$N_\mathrm{s}=5$, $m_\mathrm{max}=45\,\Msun$&-11&-8.5&-7.8&-4.5&-2.4& -7.9\\\hline
$N_\mathrm{s}=6$& -2.8&-2.0&-2.6&-1.5&-0.42&-2.0\\
\hline \hline
\multicolumn{7}{c}{Globular cluster}  \\\hline
Fiducial& -2.3&-6.8&-8.7&-5.3&-4.8&-0.83\\\hline
$\alpha=2$& 4.2&-0.50&-1.5&-1.4&-2.1&{\bf 5.8} \\\hline
$m_\mathrm{max}=30\,\Msun$& -12&-10&-8.7&-9.1&-7.0&-15\\\hline
$a_\mathrm{uni}=1$&-12&-13&-16&-8.9&-9.6&-10\\\hline
$a_\mathrm{ave}=0.3$& -0.84&-5.1&-7.8&-4.8&-5.2&-3.3\\\hline
$a_\mathrm{ave}=0.5$& -1.8&-5.0&-8.6&-4.6&-4.6&-4.2\\\hline
$a_\mathrm{ave}=0.7$& -12&-12&-14&-8.7&-7.9&-8.7\\\hline
$v_\mathrm{esc}=100\,\mathrm{km/s}$& -3.1&-7.1&-8.5&-4.8&-4.9&-1.4\\\hline
$w=0.05$& -2.1&-5.4&-7.1&-3.8&-3.9&{\bf 0.54}\\\hline
$w=0.1$&-3.3&-5.5&-7.3&-4.1&-3.5&-0.72\\\hline
$\alpha=2$, $a_\mathrm{uni}=1$& -4.1&-4.8&-5.6&-5.6&-4.8& -2.6\\\hline
$\alpha=2$, $a_\mathrm{ave}=0.3$& 4.9&{\bf 0.38}&-0.72&-1.1&-2.4& {\bf 2.6}\\\hline
$\alpha=2$, $a_\mathrm{ave}=0.5$& 0.15&-2.9&-5.0&-4.4&-4.6& -2.3
\\\hline\hline
\multicolumn{7}{c}{Field binary}  \\\hline
Fiducial& -18&-17&-19&-12&-10&-9.1
\\\hline\hline
\multicolumn{7}{c}{Migration trap}  \\\hline
Fiducial& -61&-37&-23&-17&-14&-57\\\hline
\multicolumn{7}{c}{AGN disk \citep{Tagawa20_massgap}}  \\\hline\hline
$f_\mathrm{m1g}=1$&-4.0 & -2.5 & -1.6 & -1.5 & -0.63 &-3.9\\\hline
$f_\mathrm{m1g}=1.33$&-2.7 & -1.9 & -1.5 & -0.53 & -0.19 &-2.5\\\hline
$f_\mathrm{m1g}=1.66$& -2.0 & -1.8 & -1.8 & -0.45 & -0.16 & -2.0 \\\hline
$f_\mathrm{m1g}=2$& 1.1 & {\bf 0.032} & -0.79 & -0.15 & {\bf 0.045} &{\bf 1.2}\\\hline
$f_\mathrm{m1g}=3$& 2.1 & -0.19 & -1.7 & -1.7 & -1.1 & {\bf 4.3} \\\hline
\end{tabular}
\end{center}
\end{table*}

\subsubsection{Bayes factors on spins and mass distributions}

\label{sec:results_bayesian_factor}

In the previous section we focus on the ${\bar \chi}_\mathrm{typ}$ profile, while here we use the distributions of $\chi_\mathrm{eff}$, $\chi_\mathrm{p}$, and $m_\mathrm{chirp}$ and discuss the preferred values for underlying parameters ${\bm \lambda}_0$. 

To assess the 
relative
likelihood to produce each event in different models, we calculate the 
Bayes factors
between pairs of models, 
\begin{align}
K_{\mathrm{A,B}}=\frac{P({\bm d}|A)}{P({\bm d}|B)}
\label{eq:bayesratio}
\end{align}
where
\begin{align}
\label{eq:p_da}
P({\bm d}|A)=\prod_i P({\bm d}_i|A),
\end{align}
$P({\bm d}_i|A)$ is the likelihood of obtaining data ${\bm d}_i$ observed in the GW event $i$ 
from model~$A$, 
\begin{align} 
\label{eq:p_likelihood_dia}
P({\bm d}_i|A)=&\nonumber\\\int P({\bm d}_i|m_\mathrm{chirp},\chi_\mathrm{eff},\chi_\mathrm{p})&P(m_\mathrm{chirp},\chi_\mathrm{eff},\chi_\mathrm{p}|A)\nonumber\\&dm_\mathrm{chirp}d \chi_\mathrm{eff} d\chi_\mathrm{p}\end{align}
and $P(m_\mathrm{chirp},\chi_\mathrm{eff},\chi_\mathrm{p}|A)$ is the probability distribution of $m_\mathrm{chirp}$, $\chi_\mathrm{eff}$, and $\chi_\mathrm{p}$ in model~$A$. 
We calculate the three dimensional likelihood $P({\bm d}_i|m_\mathrm{chirp},\chi_\mathrm{eff},\chi_\mathrm{p})$ 
for the events.

We calculate the Bayes factors for events with $D_\mathrm{KL}\geq D_\mathrm{KL,crit}=0$, 0.05, 0.1, 0.15, and 0.2. 
We consider $D_\mathrm{KL,crit}=0.05$ as the fiducial value, and mostly discuss the Bayes factors for $D_\mathrm{KL,crit}=0.05$ below. 
Note that the events with positive Bayes factors for $D_\mathrm{KL,crit}=0.1$, $0.15$, or $0.2$ always have positive Bayes factors also for $D_\mathrm{KL,crit}=0.05$ somewhat incidentally.

To calculate $P(m_\mathrm{chirp},\chi_\mathrm{eff},\chi_\mathrm{p}|A)$, we first count mergers in $30\times 30 \times 30$ uniform bins in $\chi_\mathrm{eff}$, $m_\mathrm{chirp}$, and $\chi_\mathrm{p}$ for model~$A$. The maximum and minimum values of $m_\mathrm{chirp}$ for the bins are set to $100$ and $5\,\Msun$, respectively. 
In this section, we generate 1000 mergers for each model. To include error distributions for 
the variables ($m_\mathrm{chirp}$, $\chi_\mathrm{eff}$, $\chi_\mathrm{p}$) to $P(m_\mathrm{chirp},\chi_\mathrm{eff},\chi_\mathrm{p}|A)$, we sample 10 different realizations for each merger event predicted by the model. 
To reduce the statistical fluctuation in the distribution of $\chi_\mathrm{eff}$, $m_\mathrm{chirp}$, and $\chi_\mathrm{p}$ due to the finite number of mergers in our models, we perform a kernel-density estimate for the distribution using Gaussian kernels whose bandwidth is chosen to satisfy the Scott's Rule \citep{Scott92}. 
We calculate $P({{\bm d}_i}|m_\mathrm{chirp},\chi_\mathrm{eff},\chi_\mathrm{p})$ by means of 300 samples generated according to 
the observed posterior distributions 
as used in the previous section.

For reference, we also calculate the Bayes factors for the two parameters, $m_\mathrm{chirp}$ and $\chi_\mathrm{eff}$, 
using the 44 events used in the analysis with $D_{\rm KL}\geq0$ in the previous section.

Table~\ref{table:bayesian_values} lists
the Bayes factors for some models relative to the fiducial model ($=B$, Table~\ref{table:parameter_fiducial}). 
The Bayes factors suggest that, 
compared to the $m_\mathrm{chirp}$, $\chi_\mathrm{eff}$, and $\chi_\mathrm{p}$ distributions typically expected for mergers in FBs and MTs (Table~\ref{table:parameters_populations}), the observed distribution is much more consistent with those in AGN disks. 
This is because high $|\chi_\mathrm{eff}|$ and low $\chi_\mathrm{p}$ expected for mergers either in FBs or MTs (panels~c and d in Figs.~\ref{fig:dist_mc_x_pop} and \ref{fig:x_mc_profile_pop}) 
are incompatible with the observed distribution of 
$|\chi_\mathrm{eff}|<\chi_\mathrm{p}$ (Fig.~\ref{fig:x_mc_profile_pop}~e).

For mergers in GCs, the models with small spin magnitudes for 1g BHs are less favored. This is presumably because infrequent hierarchical mergers ($\lesssim 20\%$) in GCs are difficult to explain typically high values of $\chi_\mathrm{p}$ if 1g BHs have low spin magnitudes. On the other hand, for $a_\mathrm{ave}=0.3$ and $\alpha=2$, the Bayes factor for $D_\mathrm{KL,crit}=0.05$ is as high as $\sim 10^{0.4}$. Thus, if mergers originate from GCs, 1g BHs are favored to have high spin magnitudes and follow a bottom heavy initial mass function. 

For mergers in AGN disks or NSCs, 
the models with 
non-zero values for initial BH spins ($a_\mathrm{ave}=0.3$) as well as a high value for $\sigma_{\chi_\mathrm{p}}(\sim 0.3$--$0.4$) have high Bayes factors of $10^{0.5}$ and $10^{0.05}$--$10^{0.4}$ for $D_\mathrm{KL,crit}=0.05$, respectively. 
This is because non-zero ${\bar \chi}_\mathrm{typ}$ at low $m_\mathrm{chirp}$ in the observed distribution (Fig.~\ref{fig_app:mc_x_obs}) can be explained by adjusting these variables (Fig.~\ref{fig:x_mc_profile_dep}). 
Also, large values 
for $\alpha$, which effectively shift the $\chi_\mathrm{typ}$ and $m_\mathrm{chirp}$ distribution toward lower $m_\mathrm{chirp}$, and accordingly raises ${\bar \chi}_\mathrm{typ}$ at low $m_\mathrm{chirp}$ 
(e.g. Fig.~\ref{fig_app:x_mc_profile_dep}~e). 
This is presumably the reason why the model with $\alpha(=2)$ has a high Bayes factor of $10^{0.8}$ at $m_\mathrm{max}\sim25\,\Msun$ compared to the models with $\alpha=1$ ($K_{A,B}\lesssim 1$). 

Preferred values for $m_\mathrm{max}$ are probably as low as $\sim 15$--$30\,\Msun$ if the typical spin magnitude for 1g BHs is low. 
For $\alpha=1$, 
in the models with $N_\mathrm{s}=3$, 4, and 5, respectively, 
$m_\mathrm{max}=25$--$30\,\Msun$, $m_\mathrm{max}=20$--$25\,\Msun$, and 
$m_\mathrm{max}=15$--$20\,\Msun$ is preferred. 
The difference in preference of $m_\mathrm{max}$ for different $N_\mathrm{s}$ is because 
both variables are constrained by the maximum $m_\mathrm{chirp}$ among the GW events. 
In any case, the preferred values of $m_\mathrm{max}=15$--$30\,\Msun$ are roughly consistent with the values estimated in the previous section.

We also compare the properties inferred from GW observations with those predicted for mergers 
in AGN disks, which are calculated from one-dimensional $N$-body simulations, combined with a semi-analytical model used in \citet{Tagawa20_massgap}. We adopt the fiducial model in \citet{Tagawa20_massgap}, while we investigate several variations in which the initial BH masses are multiplied by $f_\mathrm{m1g}=$1, 1.33, 1.66, 2, and 3 so that $m_\mathrm{max}=15$, 20, 25, 30, and 45 $\Msun$, respectively. 
Since 1g BH masses are $5$--$15\,\Msun$ in the fiducial model, the minimum BH mass is given by $5f_\mathrm{m1g}\,\Msun$, in which the minimum chirp mass is $\sim 8.7f_\mathrm{m1g}\,\Msun$. 
To eliminate a reduction of the likelihood due to the lack of 1g BHs in the low mass ranges,  
we here calculate Bayes factors only using events with $m_\mathrm{chirp}\geq 8.7f_\mathrm{m1g}\,\Msun$. 
The errors on $m_\mathrm{chirp}$, $\chi_\mathrm{eff}$, and $\chi_\mathrm{p}$ are simply given by the normal distribution with the standard deviation of $0.08\,m_\mathrm{chirp}$, $0.12$, and $0.2$, respectively. 
The Bayes factors are listed in the bottom five rows in Table~\ref{table:bayesian_values}, which indicate that $m_\mathrm{max}\sim 30\,\Msun$ ($f_\mathrm{m1g}=2$) is preferred. 
Thus, the properties predicted for AGN disk-assisted mergers are likely to be consistent with the observed properties of the GW events.

Here, events with high Bayes factors for $D_\mathrm{KL,crit}=0.05$ 
tend to have high Bayes factors for the two dimensional likelihood (bold number in the third and rightmost columns of Table~\ref{table:bayesian_values}). 
We consider that this fact would further support the preferred models discussed above.

Overall, our analyses suggest that  $m_\mathrm{max}=15$--$30\,\Msun$ with a high fraction of hierarchical mergers, or high spin magnitudes of $\sim 0.3$ for 1g BHs is favored. 
The former may support mergers in NSCs including AGN disks, while the latter may be consistent with those in GCs. 
Further events would be required to assess these possibilities in more detail.

We also discuss the spin distribution suggested in \citet{LIGO20_O3_Properties}. 
First, we compare the average and the standard deviation of $\chi_\mathrm{p}$ predicted by models and those estimated from LIGO/Virgo O1--O3a data. By analyzing the observed GW data, 
\citet{LIGO20_O3_Properties} estimated that the average and the standard deviation  of $\chi_\mathrm{p}$ are $0.21^{+0.15}_{-0.14}$ and $0.09^{+0.21}_{-0.07}$, respectively, assuming a truncated mass model. These values are consistent with models in which hierarchical mergers are frequent such as models~M1, M9--M11 (Table~\ref{table:fraction_highg}), M18--M21, M25--M28 (Table~\ref{table_app:fraction_highg}). 
Also, the average and the standard deviation $\chi_\mathrm{p}$ for the model of GC with $a_\mathrm{ave}=0.3$ is 0.25 and 0.062, respectively, which are also consistent with the values estimated from the observed data. This fact further supports our claim that frequent hierarchical mergers or high spin magnitudes of $\sim 0.3$ for 1g BHs is favored. 
Here, note that the dependence of the spins on masses expected from hierarchical mergers is taken into account in our analysis, which would be a critical difference from that in \citet{LIGO20_O3_Properties}.

Next, we discuss the fraction of mergers with positive and negative $\chi_\mathrm{eff}$. 
\citet{LIGO20_O3_Properties} analyzed the GW data observed in LIGO/Virgo O1--O3a, and estimated that $0.67^{+0.16}_{-0.16}$ (the $90\%$ credible intervals) and $0.27^{+0.17}_{-0.15}$ of mergers have $\chi_\mathrm{eff}>0.01$ and $\chi_\mathrm{eff}<-0.01$, respectively. 
In the fiducial model (Table~\ref{table:parameter_fiducial}), the fraction of mergers with $\chi_\mathrm{eff}>0.01$ is 0.54 and that for $\chi_\mathrm{eff}<-0.01$ is 0.41. 
The larger fraction for positive $\chi_\mathrm{eff}$ compared to that for negative one in the model is due to the assumed dependence of $\rho_0$ on $\chi_\mathrm{eff}$ in Eq.~\eqref{eq:rho0}. 
The fraction of mergers with negative $\chi_\mathrm{eff}$ in the model is somewhat higher than that estimated in \citet{LIGO20_O3_Properties}. 
Such difference may be due to large uncertainties for the estimated fraction, 
while it may suggest that the dependence of $\rho_0$ on $\chi_\mathrm{eff}$ is stronger than that adopted in Eq.~\eqref{eq:rho0}, or 
BH spins are moderately aligned toward the binary angular momentum directions due to interactions with gas, tidal synchronization, or alignment of spins for progenitor stars.

\begin{figure}
\begin{center}
\includegraphics[width=85mm]{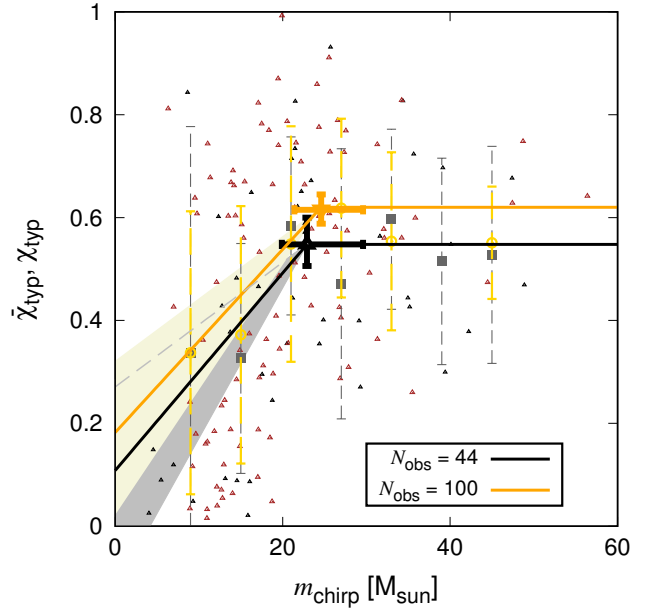}
\caption{
The ${\bar \chi}_{\rm typ}$ profile as a function of $m_\mathrm{chirp}$ constructed using the mock GW data for $100$ (orange), and $44$ (black) observed events. 
Thick lines present the recovered profiles as in Fig.~\ref{fig:x_mc_construct_obs}. 
The triangles corresponds to the median values of ${\chi}_{\rm typ}$ for all mock events. The dashed lines show $1\sigma$ credible intervals for observed values of ${\bar \chi}_{\rm typ}$ for mock events. 
}
\label{fig:x_mc_construct}
\end{center}
\end{figure}

\subsection{Reconstruction of the spin profile from mock GW data}

\label{sec:result_rec_x}

We investigate how well the ${\bar \chi}_\mathrm{typ}$ profile can be reconstructed from mock GW data ($\S\,\ref{sec:mock_data}$) for different values of $N_\mathrm{obs}$ 
by performing the MCMC method as described in $\S\,\ref{sec:method_reconstruction_spin}$.  
Fig.~\ref{fig:x_mc_construct} shows ${\bar \chi}_\mathrm{typ}$ as a function of $m_\mathrm{chirp}$ for $N_\mathrm{obs}=44$ (black) and $N_\mathrm{obs}=100$ (orange) for the model with the fiducial setting (Table~\ref{table:parameter_fiducial}) but $m_\mathrm{max}=30\,\Msun$ and $\alpha=2$, which is preferred from observed GW events ($\S\,\ref{sec:result_rec_x_obs}$ and $\S\,\ref{sec:results_bayesian_factor}$).

As the parameter estimate tends to be biased in small $N_\mathrm{obs}$, we additionally perform 10 models for $N_\mathrm{obs}=44$ with same settings with independent realizations of the initial condition. 
By averaging the estimated parameters for eleven models, 
${\bar \chi}_\mathrm{typ}$ at the plateau is $b_\mu=0.63^{+0.07}_{-0.05}$ with the standard deviation $\sigma (b_\mu)=0.04$, 
the critical chirp mass is $m_{\mathrm{crit}}=24^{+6}_{-4}\,\Msun$ with $\sigma(m_{\mathrm{crit}})=5\,\Msun$, 
and the slope of ${\bar \chi}_\mathrm{typ}$ in $m_\mathrm{chirp}<m_{\mathrm{crit}}$ is $a_\mu=25 ^{+9}_{-8}\times 10^{-3}\,\Msun^{-1}$ with $\sigma (b_\mu)=8\times 10^{-3}\,\Msun^{-1}$. 
As these uncertainties on the reconstructed parameters from the GW mock data are similar to those derived from the observed GW data in $\S\,\ref{sec:result_rec_x_obs}$, we conclude that the GW mock data are a useful tool to understand how well the spin profile can be reconstructed. 

The critical chirp mass is estimated to be 
$m_{\mathrm{crit}}=25^{+6}_{-3}\,\Msun$ for $N_\mathrm{obs}=100$, 
and $m_{\mathrm{crit}}=24.9^{+0.9}_{-0.8}\,\Msun$ for $N_\mathrm{obs}=1000$. 
Here, the estimated value of $m_{\mathrm{crit}}$ is lower than $m_\mathrm{max}$ by $\sim 20\%$ mostly because $m_\mathrm{chirp} =(m_\mathrm{1}+m_\mathrm{2})[q(1+q)^{-2}]^{3/5}\lesssim 0.87 m_\mathrm{max}$. 
As the analysis on the observed GW events in $\S\,\ref{sec:result_rec_x_obs}$ derives $m_{\mathrm{crit}}\sim 15$--$50\,\Msun$, $m_\mathrm{max}\sim 20$--$60\,\Msun$ is roughly inferred according to the relation of 
$m_\mathrm{max}\sim 1.2\,m_{\mathrm{crit}}$.

The average spin parameter ${\bar \chi}_\mathrm{typ}$ at $m_\mathrm{chirp}=m_\mathrm{min}$ is related to the typical spin magnitude of 1g BHs (e.g. Fig.~\ref{fig:x_mc_profile_dep}). 
${\bar \chi}_\mathrm{typ}$ at $m_\mathrm{chirp}=5\,\Msun$ is 
$0.20^{+0.14}_{-0.17}$ for $N_\mathrm{obs}=44$, 
$0.27^{+0.11}_{-0.13}$ for $N_\mathrm{obs}=100$, and 
$0.23^{+0.03}_{-0.02}$ for $N_\mathrm{obs}=1000$. 
These values derived from the model with $a_\mathrm{ave}=a_\mathrm{uni}=0$ are similar to the value ($\sim 0.3 \pm \sim 0.1$) derived from the observed GW data ($\S\,\ref{sec:result_rec_x_obs}$), suggesting that the typical spin magnitude of 1g BHs inferred from the observed GW events is still consistent with $\sim 0$. 

For $N_\mathrm{obs}=100$, ${\bar \chi}_\mathrm{typ}$ at the plateau is $b_\mu=0.62^{+0.04}_{-0.03}$, which is similar to the expected value for hierarchical mergers ($\sim 0.6$, $\S\,\ref{sec:result_pop_parameters}$). Also, the mass at the bending point is well constrained with $N_\mathrm{obs}=100$ as mentioned above. 
Thus, with $N_\mathrm{obs}\geq100$, 
parameters characterising properties of hierarchical mergers, e.g. a value of ${\bar \chi}_\mathrm{typ}$ and $m_\mathrm{crit}$ at the plateau, are more precisely constrained. 

Finally, 
to investigate whether the bending point is robustly verified, we also fit the distribution by a straight line, i.e. assuming $m_{\mathrm{crit}}\rightarrow \infty$ in Eq.~\eqref{eq:p_theta_lambda}, 
and calculate the Bayes factor of the model with broken lines (Eq.~\ref{eq:p_theta_lambda}) compared to the model with a single line ($m_{\mathrm{crit}}\rightarrow \infty$), where we set the likelihood function to Eq.~\eqref{eq:p_theta_lambda} with the fitted parameters. 
For $N_\mathrm{obs}=44$, 100, and 1000, the logarithm of the Bayes factor is 1.5, 2.1, and 24, respectively. 
If we adopt the Akaike information criterion \citep{Akaike74}, the model with the broken lines is preferred by a factor of $\sim 10^{1.6}$ for $N_\mathrm{obs}=100$, and the preference increases as $N_\mathrm{obs}$ increases. 
In the analysis using the observed data in $\S$~\ref{sec:result_rec_x_obs}, although we assumed the existence of the plateau, the Bayes factors using the observed events (with $D_\mathrm{KL,cri}=0$, 0.05, 0.1, and $0.15$) are in the range of $10^{-0.2}$--$10^{0.2}$, suggesting that the existence of the plateau is uncertain. 
Our analysis suggests that as the number of GW events increases to $\gtrsim O(100)$, 
the existence of the plateau can be confirmed with high significance.

\section{Summary and Conclusions}

In this paper 
we have investigated characteristic distributions of $\chi_\mathrm{eff}$, $\chi_\mathrm{p}$, $\chi_\mathrm{typ}=(\chi_\mathrm{p}^2+\chi_\mathrm{eff}^2)^{1/2}$, and $m_\mathrm{chirp}$ expected from hierarchical mergers among stellar-mass BHs. We then
used a toy model to derive 
the profile of the average of $\chi_\mathrm{typ}$ as a function of $m_\mathrm{chirp}$ for the events observed by LIGO/Virgo O1--O3a. 
We also investigated how well predictions in different models match observed spin and mass distributions by using Bayes factors. 
Finally, we estimate how well the $\chi_\mathrm{typ}$ profile can be reconstructed using mock GW data expected in hierarchical mergers. 
Our main results are summarized as follows:

\begin{enumerate}

\item 
If hierarchical mergers are frequent, and the spin distribution of first-generation (1g) BHs does not strongly depend on their mass, 
the ${\bar \chi}_\mathrm{typ}$ profile as a function of $m_\mathrm{chirp}$ is characterized by 
a monotonic increase of ${\bar \chi}_\mathrm{typ}$ with $m_\mathrm{chirp}$ 
up to
the maximum chirp mass among 1g BHs, 
and reaches a plateau of ${\bar \chi}_\mathrm{typ}$ with $\sim 0.6$ at higher $m_\mathrm{chirp}$ (Fig.~\ref{fig:x_mc_profile_dep}). 
With $\sim 50$ events, 
the plateau and the rise of $\chi_\mathrm{typ}$ to $0.6$ can be confirmed if the detection fraction of mergers of high-g BHs roughly exceeds $\sim 0.5$ and $\sim 0.15$, respectively.

\item 
The maximum mass 
for 1g BHs can be estimated by constraining the transition point between the two regimes in the ${\bar \chi}_\mathrm{typ}$ profile. 
Also, the typical spin magnitude for 1g BHs is constrained from ${\bar \chi}_\mathrm{typ}$ at around minimum $m_\mathrm{chirp}$ among GW events.

\item 
The ${\bar \chi}_\mathrm{typ}$ profile reconstructed from the LIGO/Virgo O1--O3a data prefers 
an increase in ${\bar \chi}_\mathrm{typ}$ at $m_\mathrm{chirp}\lesssim 15$--$50\,\Msun$ with $\sim 2\sigma$ confidence (Fig.~\ref{fig:x_mc_construct_obs}), 
consistent with the evolution of BH spin magnitudes by hierarchical mergers. 
The maximum mass and the typical spin magnitude of 1g BHs are loosely constrained to be $\sim 20$--$60\,\Msun$ and $\lesssim 0.4$ 
with $\sim 1\sigma$ credible intervals, respectively.

\item 
A Bayesian analysis using the $\chi_\mathrm{eff}$, $\chi_\mathrm{p}$, and $m_\mathrm{chirp}$ distributions 
suggests that 
1g BHs are preferred to have the maximum mass of $m_\mathrm{max}\sim 15$--$30\,\Msun$ 
if hierarchical mergers are frequent, 
which is consistent with mergers in AGN disks and/or nuclear star clusters. 
On the other hand, 
if mergers mainly originate from globular clusters (in which $m_\mathrm{max}$ is assumed to be $45\,\Msun$), 
1g BHs are favored to have spin magnitudes of $\sim 0.3$. 
These favored models are also consistent with the average and the standard deviation of $\chi_\mathrm{p}$ estimated in \citet{LIGO20_O3_Properties}. 

\item 
By using observed data of more than $\sim 100$ events in the future, we will be able to recover parameters characterizing the ${\bar \chi}_\mathrm{typ}$ distribution (e.g. the existence of the plateau and the value of ${\bar \chi}_\mathrm{typ}$ at the plateau $b_\mu$) more precisely.

\end{enumerate}

\acknowledgments

The authors greatly thank V. Gayathri for providing us the posterior distributions, 
Davide Gerosa for providing us values for the Kullback-Leibler divergence, 
and Berry Christopher and 
Kengo Tomida for valuable comments. 
This work is financially supported by 
the Grant-in-Aid for JSPS Research Fellowship and 
for Basic Research by the Ministry of Education, Science and Culture of Japan (HT:17H01102, 17H06360, KO:17H02869, 17H01102, 17H06360). 
ZH acknowledges support from NASA grant NNX15AB19G and NSF grants 1715661 and 2006176.
This work received founding from the European Research Council (ERC) under the European Union's Horizon 2020 Programme for Research and Innovation ERC-2014-STG under grant agreement No. 638435 (GalNUC) (to BK). 
IB acknowledges support from the Alfred P. Sloan Foundation and from the University of Florida. 
Simulations and analyses were carried out on Cray XC50 
at the Center for Computational Astrophysics, National Astronomical Observatory of Japan.

\section*{DATA AVAILABILITY}

The data underlying this article will be shared on reasonable
request to the corresponding author.

\bibliographystyle{yahapj.bst}
\bibliography{agn_bhm}

\begin{thebibliography}{}
\providecommand\natexlab[1]{#1}
\providecommand\JournalTitle[1]{#1}

\bibitem[{{Aasi} {et~al.}(2015)}]{2015CQGra..32g4001L}
{Aasi}, J., {et~al.} 2015,
  \href{http://dx.doi.org/10.1088/0264-9381/32/7/074001}{\JournalTitle{CQG},
  32, 074001}

\bibitem[{{Abbott} {et~al.}(2019){Abbott}, {Abbott}, {Abbott}, {Abraham},
  {Acernese}, {Ackley}, {Adams}, {Adhikari}, {Adya}, {Affeldt}, {Agathos},
  {Agatsuma}, {Aggarwal}, {Aguiar}, {Aiello}, {Ain}, {Ajith}, {Allen},
  {Allocca}, {Aloy}, {Altin}, {Amato}, {Ananyeva}, {Anderson}, {LIGO Scientific
  Collaboration}, \& {Virgo Collaboration}}]{TheLIGO18}
{Abbott}, B.~P., {Abbott}, R., {Abbott}, T.~D., {et~al.} 2019,
  \href{http://dx.doi.org/10.1103/PhysRevX.9.031040}{\JournalTitle{Physical
  Review X}, 9, 031040}

\bibitem[{{Abbott} {et~al.}(2020{\natexlab{a}}){Abbott}, {Abbott}, {Abraham},
  {Acernese}, {Ackley}, {Adams}, {Adhikari}, {Adya}, {Affeldt}, {Agathos},
  {Agatsuma}, {Aggarwal}, {Aguiar}, {Aich}, {Aiello}, {Ain}, {Ajith}, {LIGO
  Scientific Collaboration}, \& {Virgo Collaboration}}]{LIGO20_GW190814}
{Abbott}, R., {Abbott}, T.~D., {Abraham}, S., {et~al.} 2020{\natexlab{a}},
  \href{http://dx.doi.org/10.3847/2041-8213/ab960f}{\JournalTitle{\apjl}, 896,
  L44}

\bibitem[{{Abbott} {et~al.}(2020{\natexlab{b}}){Abbott}, {Abbott}, {Abraham},
  {Acernese}, {Ackley}, {Adams}, {Adams}, {Adhikari}, {Adya}, {Affeldt},
  {Agathos}, {Agatsuma}, {Aggarwal}, {Aguiar}, {Aiello}, {Ain}, \&
  {Ajith}}]{LIGO20_O3_Catalog}
---. 2020{\natexlab{b}}, \JournalTitle{arXiv e-prints}, arXiv:2010.14527

\bibitem[{{Abbott} {et~al.}(2020{\natexlab{c}}){Abbott}, {Abbott}, {Abraham},
  {Acernese}, {Ackley}, {Adams}, {Adhikari}, {Adya}, {Affeldt}, {Agathos},
  {Agatsuma}, {Aggarwal}, {Aguiar}, {Aich}, {Aiello}, {Ain}, {Ajith}, {Akcay},
  {Allen}, {Allocca}, {Altin}, {Amato}, {Anand}, {Ananyeva}, {Anderson}, {LIGO
  Scientific Collaboration}, \& {Virgo Collaboration}}]{LIGO20_GW190521_astro}
---. 2020{\natexlab{c}},
  \href{http://dx.doi.org/10.3847/2041-8213/aba493}{\JournalTitle{\apjl}, 900,
  L13}

\bibitem[{{Acernese} {et~al.}(2015)}]{2015CQGra..32b4001A}
{Acernese}, F., {et~al.} 2015,
  \href{http://dx.doi.org/10.1088/0264-9381/32/2/024001}{\JournalTitle{CQG},
  32, 024001}

\bibitem[{{Akaike}(1974)}]{Akaike74}
{Akaike}, H. 1974,
  \href{http://dx.doi.org/10.1109/TAC.1974.1100705}{\JournalTitle{IEEE
  Transactions on Automatic Control}, 19, 716}

\bibitem[{{Antonini} {et~al.}(2019){Antonini}, {Gieles}, \&
  {Gualandris}}]{Antonini19}
{Antonini}, F., {Gieles}, M., \& {Gualandris}, A. 2019,
  \href{http://dx.doi.org/10.1093/mnras/stz1149}{\JournalTitle{\mnras}, 486,
  5008}

\bibitem[{{Antonini} {et~al.}(2017){Antonini}, {Toonen}, \&
  {Hamers}}]{Antonini17}
{Antonini}, F., {Toonen}, S., \& {Hamers}, A.~S. 2017,
  \href{http://dx.doi.org/10.3847/1538-4357/aa6f5e}{\JournalTitle{\apj}, 841,
  77}

\bibitem[{Apostolatos {et~al.}(1994)Apostolatos, Cutler, Sussman, \&
  Thorne}]{Apostolatos94}
Apostolatos, T.~A., Cutler, C., Sussman, G.~J., \& Thorne, K.~S. 1994,
  \href{http://dx.doi.org/10.1103/PhysRevD.49.6274}{\JournalTitle{Phys. Rev.
  D}, 49, 6274}

\bibitem[{{Arca Sedda}(2020)}]{ArcaSedda20}
{Arca Sedda}, M. 2020, \JournalTitle{arXiv e-prints}, arXiv:2002.04037

\bibitem[{{Askar} {et~al.}(2020){Askar}, {Davies}, \& {Church}}]{Askar20}
{Askar}, A., {Davies}, M.~B., \& {Church}, R.~P. 2020, \JournalTitle{arXiv
  e-prints}, arXiv:2006.04922

\bibitem[{{Banerjee}(2017)}]{Banerjee17}
{Banerjee}, S. 2017,
  \href{http://adsabs.harvard.edu/abs/2017MNRAS.467..524B}{\JournalTitle{\mnras},
  467, 524}

\bibitem[{{Bardeen} \& {Petterson}(1975)}]{Bardeen75}
{Bardeen}, J.~M., \& {Petterson}, J.~A. 1975,
  \href{http://adsabs.harvard.edu/abs/1975ApJ...195L..65B}{\JournalTitle{\apj},
  195, L65}

\bibitem[{{Bartos} {et~al.}(2017){Bartos}, {Kocsis}, {Haiman}, \&
  {M{\'a}rka}}]{Bartos17}
{Bartos}, I., {Kocsis}, B., {Haiman}, Z., \& {M{\'a}rka}, S. 2017,
  \href{http://adsabs.harvard.edu/abs/2017ApJ...835..165B}{\JournalTitle{\apj},
  835, 165}

\bibitem[{{Bavera} {et~al.}(2019){Bavera}, {Fragos}, {Qin}, {Zapartas},
  {Neijssel}, {Mandel}, {Batta}, {Gaebel}, {Kimball}, \&
  {Stevenson}}]{Bavera19}
{Bavera}, S.~S., {Fragos}, T., {Qin}, Y., {et~al.} 2019, \JournalTitle{arXiv
  e-prints}, arXiv:1906.12257

\bibitem[{{Belczynski} {et~al.}(2016){Belczynski}, {Daniel}, {Bulik}, \&
  {O'Shaughnessy}}]{Belczynski16}
{Belczynski}, K., {Daniel}, E.~H., {Bulik}, T., \& {O'Shaughnessy}, R. 2016,
  \href{http://adsabs.harvard.edu/abs/2016Natur.534..512B}{\JournalTitle{\nat},
  534, 512}

\bibitem[{{Bellovary} {et~al.}(2016){Bellovary}, {Mac Low}, {McKernan}, \&
  {Ford}}]{Bellovary16}
{Bellovary}, J.~M., {Mac Low}, M.-M., {McKernan}, B., \& {Ford}, K.~E.~S. 2016,
  \href{http://dx.doi.org/10.3847/2041-8205/819/2/L17}{\JournalTitle{\apjl},
  819, L17}

\bibitem[{{Brodie} {et~al.}(2014){Brodie}, {Romanowsky}, {Strader}, {Forbes},
  {Foster}, {Jennings}, {Pastorello}, {Pota}, {Usher}, {Blom}, {Kader},
  {Roediger}, {Spitler}, {Villaume}, {Arnold}, {Kartha}, \&
  {Woodley}}]{Brodie14}
{Brodie}, J.~P., {Romanowsky}, A.~J., {Strader}, J., {et~al.} 2014,
  \href{http://dx.doi.org/10.1088/0004-637X/796/1/52}{\JournalTitle{\apj}, 796,
  52}

\bibitem[{{Buonanno} {et~al.}(2008){Buonanno}, {Kidder}, \&
  {Lehner}}]{Buonanno08}
{Buonanno}, A., {Kidder}, L.~E., \& {Lehner}, L. 2008,
  \href{http://dx.doi.org/10.1103/PhysRevD.77.026004}{\JournalTitle{\prd}, 77,
  026004}

\bibitem[{{Chatzopoulos} \& {Wheeler}(2012)}]{Chatzopoulos12}
{Chatzopoulos}, E., \& {Wheeler}, J.~C. 2012,
  \href{http://dx.doi.org/10.1088/0004-637X/748/1/42}{\JournalTitle{\apj}, 748,
  42}

\bibitem[{{Chen} {et~al.}(2017){Chen}, {Holz}, {Miller}, {Evans}, {Vitale}, \&
  {Creighton}}]{Chen17}
{Chen}, H.-Y., {Holz}, D.~E., {Miller}, J., {et~al.} 2017, \JournalTitle{arXiv
  e-prints}, arXiv:1709.08079

\bibitem[{{de Mink} \& {Mandel}(2016)}]{deMink16}
{de Mink}, S.~E., \& {Mandel}, I. 2016,
  \href{http://adsabs.harvard.edu/abs/2016MNRAS.460.3545D}{\JournalTitle{\mnras},
  460, 3545}

\bibitem[{{Di Carlo} {et~al.}(2019){Di Carlo}, {Giacobbo}, {Mapelli},
  {Pasquato}, {Spera}, {Wang}, \& {Haardt}}]{DiCarlo19}
{Di Carlo}, U.~N., {Giacobbo}, N., {Mapelli}, M., {et~al.} 2019,
  \href{http://dx.doi.org/10.1093/mnras/stz1453}{\JournalTitle{\mnras}, 487,
  2947}

\bibitem[{{Do} {et~al.}(2018){Do}, {Kerzendorf}, {Konopacky}, {Marcinik},
  {Ghez}, {Lu}, \& {Morris}}]{Do18}
{Do}, T., {Kerzendorf}, W., {Konopacky}, Q., {et~al.} 2018,
  \href{http://dx.doi.org/10.3847/2041-8213/aaaec3}{\JournalTitle{\apjl}, 855,
  L5}

\bibitem[{{Doctor} {et~al.}(2020){Doctor}, {Wysocki}, {O'Shaughnessy}, {Holz},
  \& {Farr}}]{Doctor20}
{Doctor}, Z., {Wysocki}, D., {O'Shaughnessy}, R., {Holz}, D.~E., \& {Farr}, B.
  2020, \href{http://dx.doi.org/10.3847/1538-4357/ab7fac}{\JournalTitle{\apj},
  893, 35}

\bibitem[{{Dominik} {et~al.}(2012){Dominik}, {Belczynski}, {Fryer}, {Holz},
  {Berti}, {Bulik}, {Mandel}, \& {O'Shughnessy}}]{Dominik12}
{Dominik}, M., {Belczynski}, K., {Fryer}, C., {et~al.} 2012,
  \href{http://adsabs.harvard.edu/abs/2012ApJ...759...52D}{\JournalTitle{\apj},
  759, 52}

\bibitem[{{Farmer} {et~al.}(2019){Farmer}, {Renzo}, {de Mink}, {Marchant}, \&
  {Justham}}]{Farmer19}
{Farmer}, R., {Renzo}, M., {de Mink}, S.~E., {Marchant}, P., \& {Justham}, S.
  2019, \href{http://dx.doi.org/10.3847/1538-4357/ab518b}{\JournalTitle{\apj},
  887, 53}

\bibitem[{{Fishbach} \& {Holz}(2020)}]{Fishbach20}
{Fishbach}, M., \& {Holz}, D.~E. 2020,
  \href{http://dx.doi.org/10.3847/2041-8213/ab7247}{\JournalTitle{\apjl}, 891,
  L27}

\bibitem[{{Fishbach} {et~al.}(2017){Fishbach}, {Holz}, \&
  {Farr}}]{Fishbach17_hierarchical}
{Fishbach}, M., {Holz}, D.~E., \& {Farr}, B. 2017,
  \href{http://dx.doi.org/10.3847/2041-8213/aa7045}{\JournalTitle{\apjl}, 840,
  L24}

\bibitem[{{Fishbach} {et~al.}(2018){Fishbach}, {Holz}, \& {Farr}}]{Fishbach18}
{Fishbach}, M., {Holz}, D.~E., \& {Farr}, W.~M. 2018,
  \href{http://dx.doi.org/10.3847/2041-8213/aad800}{\JournalTitle{\apjl}, 863,
  L41}

\bibitem[{{Fragione} {et~al.}(2019){Fragione}, {Grishin}, {Leigh}, {Perets}, \&
  {Perna}}]{Fragione19b_GN}
{Fragione}, G., {Grishin}, E., {Leigh}, N. W.~C., {Perets}, H.~B., \& {Perna},
  R. 2019,
  \href{http://dx.doi.org/10.1093/mnras/stz1651}{\JournalTitle{\mnras}, 488,
  47}

\bibitem[{{Fragione} \& {Kocsis}(2018)}]{Fragione18_EvolvingGC}
{Fragione}, G., \& {Kocsis}, B. 2018,
  \href{http://dx.doi.org/10.1103/PhysRevLett.121.161103}{\JournalTitle{\prl},
  121, 161103}

\bibitem[{{Fragione} \& {Kocsis}(2019)}]{Fragione19}
---. 2019,
  \href{http://dx.doi.org/10.1093/mnras/stz1175}{\JournalTitle{\mnras}, 486,
  4781}

\bibitem[{{Fragione} {et~al.}(2020){Fragione}, {Loeb}, \&
  {Rasio}}]{Fragione2020_Retantion}
{Fragione}, G., {Loeb}, A., \& {Rasio}, F.~A. 2020,
  \href{http://dx.doi.org/10.3847/2041-8213/abbc0a}{\JournalTitle{\apjl}, 902,
  L26}

\bibitem[{{Fuller} \& {Ma}(2019)}]{Fuller19_massive}
{Fuller}, J., \& {Ma}, L. 2019,
  \href{http://dx.doi.org/10.3847/2041-8213/ab339b}{\JournalTitle{\apjl}, 881,
  L1}

\bibitem[{{Gerosa} \& {Berti}(2017)}]{Gerosa17}
{Gerosa}, D., \& {Berti}, E. 2017,
  \href{http://dx.doi.org/10.1103/PhysRevD.95.124046}{\JournalTitle{\prd}, 95,
  124046}

\bibitem[{{Gerosa} {et~al.}(2019){Gerosa}, {Lima}, {Berti}, {Sperhake},
  {Kesden}, \& {O{\textquoteright}Shaughnessy}}]{Gerosa19}
{Gerosa}, D., {Lima}, A., {Berti}, E., {et~al.} 2019,
  \href{http://dx.doi.org/10.1088/1361-6382/ab14ae}{\JournalTitle{Classical and
  Quantum Gravity}, 36, 105003}

\bibitem[{{Gerosa} {et~al.}(2020{\natexlab{a}}){Gerosa}, {Mould}, {Gangardt},
  {Schmidt}, {Pratten}, \& {Thomas}}]{Gerosa20_xp}
{Gerosa}, D., {Mould}, M., {Gangardt}, D., {et~al.} 2020{\natexlab{a}},
  \JournalTitle{arXiv e-prints}, arXiv:2011.11948

\bibitem[{{Gerosa} {et~al.}(2020{\natexlab{b}}){Gerosa}, {Vitale}, \&
  {Berti}}]{Gerosa20}
{Gerosa}, D., {Vitale}, S., \& {Berti}, E. 2020{\natexlab{b}},
  \JournalTitle{arXiv e-prints}, arXiv:2005.04243

\bibitem[{{Gond{\'a}n} {et~al.}(2018){Gond{\'a}n}, {Kocsis}, {Raffai}, \&
  {Frei}}]{Gondan18a}
{Gond{\'a}n}, L., {Kocsis}, B., {Raffai}, P., \& {Frei}, Z. 2018,
  \href{http://dx.doi.org/10.3847/1538-4357/aabfee}{\JournalTitle{\apj}, 860,
  5}

\bibitem[{{Hamers} \& {Safarzadeh}(2020)}]{Hamers20_GW190412}
{Hamers}, A.~S., \& {Safarzadeh}, M. 2020,
  \href{http://dx.doi.org/10.3847/1538-4357/ab9b27}{\JournalTitle{\apj}, 898,
  99}

\bibitem[{{Hannam} {et~al.}(2014){Hannam}, {Schmidt}, {Boh{\'e}}, {Haegel},
  {Husa}, {Ohme}, {Pratten}, \& {P{\"u}rrer}}]{Hannam2014}
{Hannam}, M., {Schmidt}, P., {Boh{\'e}}, A., {et~al.} 2014,
  \href{http://dx.doi.org/10.1103/PhysRevLett.113.151101}{\JournalTitle{\prl},
  113, 151101}

\bibitem[{Hastings(1970)}]{Hastings1970}
Hastings, W.~K. 1970,
  \href{http://dx.doi.org/10.1093/biomet/57.1.97}{\JournalTitle{Biometrika},
  57, 97}

\bibitem[{{Hotokezaka} \& {Piran}(2017)}]{Hotokezaka17}
{Hotokezaka}, K., \& {Piran}, T. 2017,
  \href{http://dx.doi.org/10.3847/1538-4357/aa6f61}{\JournalTitle{\apj}, 842,
  111}

\bibitem[{{Inayoshi} {et~al.}(2017){Inayoshi}, {Hirai}, {Kinugawa}, \&
  {Hotokezaka}}]{Inayoshi17}
{Inayoshi}, K., {Hirai}, R., {Kinugawa}, T., \& {Hotokezaka}, K. 2017,
  \href{http://adsabs.harvard.edu/abs/2017MNRAS.468.5020I}{\JournalTitle{\mnras},
  468, 5020}

\bibitem[{{Ivanova} {et~al.}(2013){Ivanova}, {Justham}, {Chen},
  {et~al.}}]{Ivanova13}
{Ivanova}, N., {Justham}, S., {Chen}, X., {et~al.} 2013,
  \href{http://adsabs.harvard.edu/abs/2013A%26ARv..21...59I}{\JournalTitle{The
  Astronomy and Astrophysics Review}, 21, 59}

\bibitem[{{Kalogera}(2000)}]{Kalogera2000}
{Kalogera}, V. 2000,
  \href{http://dx.doi.org/10.1086/309400}{\JournalTitle{\apj}, 541, 319}

\bibitem[{{Kesden} {et~al.}(2010){Kesden}, {Sperhake}, \& {Berti}}]{Kesden10}
{Kesden}, M., {Sperhake}, U., \& {Berti}, E. 2010,
  \href{http://dx.doi.org/10.1103/PhysRevD.81.084054}{\JournalTitle{\prd}, 81,
  084054}

\bibitem[{Kidder(1995)}]{Kidder95}
Kidder, L.~E. 1995,
  \href{http://dx.doi.org/10.1103/PhysRevD.52.821}{\JournalTitle{Phys. Rev. D},
  52, 821}

\bibitem[{{Kimball} {et~al.}(2020{\natexlab{a}}){Kimball}, {Talbot}, {Berry},
  {Carney}, {Zevin}, {Thrane}, \& {Kalogera}}]{Kimball20}
{Kimball}, C., {Talbot}, C., {Berry}, C. P.~L., {et~al.} 2020{\natexlab{a}},
  \href{http://dx.doi.org/10.3847/1538-4357/aba518}{\JournalTitle{\apj}, 900,
  177}

\bibitem[{{Kimball} {et~al.}(2020{\natexlab{b}}){Kimball}, {Talbot}, {Berry},
  {Zevin}, {Thrane}, {Kalogera}, {Buscicchio}, {Carney}, {Dent}, {Middleton},
  {Payne}, {Veitch}, \& {Williams}}]{Kimball2020b}
---. 2020{\natexlab{b}}, \JournalTitle{arXiv e-prints}, arXiv:2011.05332

\bibitem[{{Kinugawa} {et~al.}(2014){Kinugawa}, {Inayoshi}, {Hotokezaka},
  {Nakauchi}, \& T.}]{Kinugawa14}
{Kinugawa}, T., {Inayoshi}, K., {Hotokezaka}, K., {Nakauchi}, D., \& T., N.
  2014,
  \href{http://adsabs.harvard.edu/abs/2014MNRAS.442.2963K}{\JournalTitle{\mnras},
  442, 2963}

\bibitem[{{Kocsis} {et~al.}(2011){Kocsis}, {Yunes}, \& {Loeb}}]{Kocsis11}
{Kocsis}, B., {Yunes}, N., \& {Loeb}, A. 2011,
  \href{http://dx.doi.org/10.1103/PhysRevD.84.024032}{\JournalTitle{\prd}, 84,
  024032}

\bibitem[{{Kumamoto} {et~al.}(2018){Kumamoto}, {Fujii}, \&
  {Tanikawa}}]{Kumamoto18}
{Kumamoto}, J., {Fujii}, M.~S., \& {Tanikawa}, A. 2018,
  \href{http://adsabs.harvard.edu/abs/2018arXiv181106726K}{\JournalTitle{arXiv
  e-prints}}, \href{http://arxiv.org/abs/1811.06726}{{\sffamily
  arXiv:1811.06726}}

\bibitem[{{Leaman} {et~al.}(2013){Leaman}, {VandenBerg}, \&
  {Mendel}}]{Leaman13}
{Leaman}, R., {VandenBerg}, D.~A., \& {Mendel}, J.~T. 2013,
  \href{http://dx.doi.org/10.1093/mnras/stt1540}{\JournalTitle{\mnras}, 436,
  122}

\bibitem[{{LIGO Scientific Collaboration} \& {Virgo
  Collaboration}(2020)}]{LIGO20_O12_HP}
{LIGO Scientific Collaboration}, \& {Virgo Collaboration}. 2020,
  \href{https://dcc.ligo.org/LIGO-P1800370/public}{\JournalTitle{LIGO Document
  P1800370-v5 https://dcc.ligo.org/LIGO-P1800370/public}}

\bibitem[{{LIGO Scientific Collaboration} \& {Virgo
  Collaboration}(2021)}]{LIGO21_O3_HP}
---. 2021, \href{https://dcc.ligo.org/LIGO-P2000223/public/}{\JournalTitle{LIGO
  Document P2000223-v7, https://dcc.ligo.org/LIGO-P2000223/public/}}

\bibitem[{{Liu} \& {Lai}(2020)}]{LiuLai20_GWevents}
{Liu}, B., \& {Lai}, D. 2020, \JournalTitle{arXiv e-prints}, arXiv:2009.10068

\bibitem[{{Lubow} {et~al.}(1999){Lubow}, {Seibert}, \& {Artymowicz}}]{Lubow99}
{Lubow}, S.~H., {Seibert}, M., \& {Artymowicz}, P. 1999,
  \href{http://dx.doi.org/10.1086/308045}{\JournalTitle{\apj}, 526, 1001}

\bibitem[{{Mandel} \& {de Mink}(2016)}]{Mandel16}
{Mandel}, I., \& {de Mink}, S.~E. 2016,
  \href{http://adsabs.harvard.edu/abs/2016MNRAS.458.2634M}{\JournalTitle{\mnras},
  458, 2634}

\bibitem[{{Mandel} {et~al.}(2019){Mandel}, {Farr}, \& {Gair}}]{Mandel19}
{Mandel}, I., {Farr}, W.~M., \& {Gair}, J.~R. 2019,
  \href{http://dx.doi.org/10.1093/mnras/stz896}{\JournalTitle{\mnras}, 486,
  1086}

\bibitem[{{Mapelli} {et~al.}(2020){Mapelli}, {Santoliquido}, {Bouffanais},
  {Arca Sedda}, {Giacobbo}, {Artale}, \& {Ballone}}]{Mapelli20}
{Mapelli}, M., {Santoliquido}, F., {Bouffanais}, Y., {et~al.} 2020,
  \JournalTitle{arXiv e-prints}, arXiv:2007.15022

\bibitem[{{Marchant} {et~al.}(2016){Marchant}, {Langer}, {Podsiadlowski},
  {Tauris}, \& {Moriya}}]{Marchant16}
{Marchant}, P., {Langer}, N., {Podsiadlowski}, P., {Tauris}, T., \& {Moriya},
  T. 2016,
  \href{http://adsabs.harvard.edu/abs/2016A%26A...588A..50M}{\JournalTitle{A\&A},
  588, A50}

\bibitem[{{McKernan} {et~al.}(2020{\natexlab{a}}){McKernan}, {Ford}, \&
  {O'Shaughnessy}}]{McKernan20_NSWD}
{McKernan}, B., {Ford}, K.~E.~S., \& {O'Shaughnessy}, R. 2020{\natexlab{a}},
  \href{http://dx.doi.org/10.1093/mnras/staa2681}{\JournalTitle{\mnras}, 498,
  4088}

\bibitem[{{McKernan} {et~al.}(2020{\natexlab{b}}){McKernan}, {Ford},
  {O'Shaugnessy}, \& {Wysocki}}]{McKernan19}
{McKernan}, B., {Ford}, K.~E.~S., {O'Shaugnessy}, R., \& {Wysocki}, D.
  2020{\natexlab{b}},
  \href{http://dx.doi.org/10.1093/mnras/staa740}{\JournalTitle{\mnras}, 494,
  1203}

\bibitem[{{McKernan} {et~al.}(2018){McKernan}, {Ford}, {Bellovary}, {Leigh},
  {Haiman}, {Kocsis}, {Lyra}, {Mac Low}, {Metzger}, {O'Dowd}, {Endlich}, \&
  {Rosen}}]{McKernan17}
{McKernan}, B., {Ford}, K.~E.~S., {Bellovary}, J., {et~al.} 2018,
  \href{http://dx.doi.org/10.3847/1538-4357/aadae5}{\JournalTitle{\apj}, 866,
  66}

\bibitem[{{Michaely} \& {Perets}(2019)}]{Michaely19}
{Michaely}, E., \& {Perets}, H.~B. 2019,
  \href{http://dx.doi.org/10.3847/2041-8213/ab5b9b}{\JournalTitle{\apjl}, 887,
  L36}

\bibitem[{{Moody} {et~al.}(2019){Moody}, {Shi}, \& {Stone}}]{Moody19}
{Moody}, M.~S.~L., {Shi}, J.-M., \& {Stone}, J.~M. 2019,
  \href{http://adsabs.harvard.edu/abs/2019arXiv190300008M}{\JournalTitle{arXiv
  e-prints}}, \href{http://arxiv.org/abs/1903.00008}{{\sffamily
  arXiv:1903.00008}}

\bibitem[{{O'Leary} {et~al.}(2009){O'Leary}, {Kocsis}, \& {Loeb}}]{OLeary09}
{O'Leary}, R.~M., {Kocsis}, B., \& {Loeb}, A. 2009,
  \href{http://adsabs.harvard.edu/abs/2009MNRAS.395.2127O}{\JournalTitle{\mnras},
  395, 2127}

\bibitem[{{O'Leary} {et~al.}(2016){O'Leary}, {Meiron}, \& {Kocsis}}]{OLeary16}
{O'Leary}, R.~M., {Meiron}, Y., \& {Kocsis}, B. 2016,
  \href{http://adsabs.harvard.edu/abs/2016ApJ...824L..12O}{\JournalTitle{ApJL},
  824, L12}

\bibitem[{{Olejak} {et~al.}(2020){Olejak}, {Fishbach}, {Belczynski}, {Holz},
  {Lasota}, {Miller}, \& {Bulik}}]{Olejak20_GW190412}
{Olejak}, A., {Fishbach}, M., {Belczynski}, K., {et~al.} 2020,
  \href{http://dx.doi.org/10.3847/2041-8213/abb5b5}{\JournalTitle{\apjl}, 901,
  L39}

\bibitem[{{Paczynski}(1976)}]{Paczynski76}
{Paczynski}, B. 1976,
  \href{http://adsabs.harvard.edu/abs/1976IAUS...73...75P}{\JournalTitle{in IAU
  Symposium, Structure and Evolution of Close Binary Systems}, 73, 75}

\bibitem[{{Pan} \& {Yang}(2021)}]{Pan2021}
{Pan}, Z., \& {Yang}, H. 2021, \JournalTitle{arXiv e-prints}, arXiv:2101.09146

\bibitem[{{Pavlovskii} {et~al.}(2017){Pavlovskii}, {Ivanova}, {Belczynski}, \&
  {Van}}]{Pavlovskii17}
{Pavlovskii}, K., {Ivanova}, N., {Belczynski}, K., \& {Van}, K.~X. 2017,
  \href{http://adsabs.harvard.edu/abs/2017MNRAS.465.2092P}{\JournalTitle{\mnras},
  465, 2092}

\bibitem[{{Peng} {et~al.}(2006){Peng}, {Jord{\'a}n}, {C{\^o}t{\'e}},
  {Blakeslee}, {Ferrarese}, {Mei}, {West}, {Merritt}, {Milosavljevi{\'c}}, \&
  {Tonry}}]{Peng06}
{Peng}, E.~W., {Jord{\'a}n}, A., {C{\^o}t{\'e}}, P., {et~al.} 2006,
  \href{http://dx.doi.org/10.1086/498210}{\JournalTitle{\apj}, 639, 95}

\bibitem[{{Planck Collaboration} {et~al.}(2016){Planck Collaboration}, {Ade},
  {Aghanim}, {Arnaud}, {Ashdown}, {Aumont}, {Baccigalupi}, {Banday},
  {Barreiro}, {Bartlett}, {Bartolo}, {Battaner}, {Battye}, {Benabed},
  {Beno{\^\i}t}, {Benoit-L{\'e}vy}, {Bernard}, {Bersanelli}, {Bielewicz},
  {Bock}, {Bonaldi}, {Bonavera}, {Bond}, {Borrill}, {Bouchet}, {Boulanger},
  {Bucher}, {Burigana}, {Butler}, {Calabrese}, {Cardoso}, {Catalano},
  {Challinor}, {Chamballu}, {Chary}, {Chiang}, {Chluba}, {Christensen},
  {Church}, {Clements}, {Colombi}, {Colombo}, {Combet}, {Coulais}, {Crill},
  {Curto}, {Cuttaia}, {Danese}, {Davies}, {Davis}, {de Bernardis}, {de Rosa},
  {de Zotti}, {Delabrouille}, {D{\'e}sert}, {Di Valentino}, {Dickinson},
  {Diego}, {Dolag}, {Dole}, {Donzelli}, {Dor{\'e}}, {Douspis}, {Ducout},
  {Dunkley}, {Dupac}, {Efstathiou}, {Elsner}, {En{\ss}lin}, {Eriksen},
  {Farhang}, {Fergusson}, {Finelli}, {Forni}, {Frailis}, {Fraisse},
  {Franceschi}, {Frejsel}, {Galeotta}, {Galli}, {Ganga}, {Gauthier}, {Gerbino},
  {Ghosh}, {Giard}, {Giraud-H{\'e}raud}, {Giusarma}, {Gjerl{\o}w},
  {Gonz{\'a}lez-Nuevo}, {G{\'o}rski}, {Gratton}, {Gregorio}, {Gruppuso},
  {Gudmundsson}, {Hamann}, {Hansen}, {Hanson}, {Harrison}, {Helou},
  {Henrot-Versill{\'e}}, {Hern{\'a}ndez-Monteagudo}, {Herranz}, {Hildebrand t},
  {Hivon}, {Hobson}, {Holmes}, {Hornstrup}, {Hovest}, {Huang}, {Huffenberger},
  {Hurier}, {Jaffe}, {Jaffe}, {Jones}, {Juvela}, {Keih{\"a}nen}, {Keskitalo},
  {Kisner}, {Kneissl}, {Knoche}, {Knox}, {Kunz}, {Kurki-Suonio}, {Lagache},
  {L{\"a}hteenm{\"a}ki}, {Lamarre}, {Lasenby}, {Lattanzi}, {Lawrence}, {Leahy},
  {Leonardi}, {Lesgourgues}, {Levrier}, {Lewis}, {Liguori}, {Lilje},
  {Linden-V{\o}rnle}, {L{\'o}pez-Caniego}, {Lubin}, {Mac{\'\i}as-P{\'e}rez},
  {Maggio}, {Maino}, {Mandolesi}, {Mangilli}, {Marchini}, {Maris}, {Martin},
  {Martinelli}, {Mart{\'\i}nez-Gonz{\'a}lez}, {Masi}, {Matarrese}, {McGehee},
  {Meinhold}, {Melchiorri}, {Melin}, {Mendes}, {Mennella}, {Migliaccio},
  {Millea}, {Mitra}, {Miville-Desch{\^e}nes}, {Moneti}, {Montier}, {Morgante},
  {Mortlock}, {Moss}, {Munshi}, {Murphy}, {Naselsky}, {Nati}, {Natoli},
  {Netterfield}, {N{\o}rgaard-Nielsen}, {Noviello}, {Novikov}, {Novikov},
  {Oxborrow}, {Paci}, {Pagano}, {Pajot}, {Paladini}, {Paoletti}, {Partridge},
  {Pasian}, {Patanchon}, {Pearson}, {Perdereau}, {Perotto}, {Perrotta},
  {Pettorino}, {Piacentini}, {Piat}, {Pierpaoli}, {Pietrobon}, {Plaszczynski},
  {Pointecouteau}, {Polenta}, {Popa}, {Pratt}, {Pr{\'e}zeau}, {Prunet},
  {Puget}, {Rachen}, {Reach}, {Rebolo}, {Reinecke}, {Remazeilles}, {Renault},
  {Renzi}, {Ristorcelli}, {Rocha}, {Rosset}, {Rossetti}, {Roudier},
  {Rouill{\'e} d'Orfeuil}, {Rowan-Robinson}, {Rubi{\~n}o-Mart{\'\i}n},
  {Rusholme}, {Said}, {Salvatelli}, {Salvati}, {Sandri}, {Santos},
  {Savelainen}, {Savini}, {Scott}, {Seiffert}, {Serra}, {Shellard}, {Spencer},
  {Spinelli}, {Stolyarov}, {Stompor}, {Sudiwala}, {Sunyaev}, {Sutton},
  {Suur-Uski}, {Sygnet}, {Tauber}, {Terenzi}, {Toffolatti}, {Tomasi},
  {Tristram}, {Trombetti}, {Tucci}, {Tuovinen}, {T{\"u}rler}, {Umana},
  {Valenziano}, {Valiviita}, {Van Tent}, {Vielva}, {Villa}, {Wade}, {Wandelt},
  {Wehus}, {White}, {White}, {Wilkinson}, {Yvon}, {Zacchei}, \&
  {Zonca}}]{Plank16}
{Planck Collaboration}, {Ade}, P.~A.~R., {Aghanim}, N., {et~al.} 2016,
  \href{http://dx.doi.org/10.1051/0004-6361/201525830}{\JournalTitle{\aap},
  594, A13}

\bibitem[{{Portegies Zwart} \& {McMillan}(2000)}]{PortegiesZwart00}
{Portegies Zwart}, S.~F., \& {McMillan}, S.~L.~W. 2000,
  \href{http://adsabs.harvard.edu/abs/2000ApJ...528L..17P}{\JournalTitle{\apj},
  528, L17}

\bibitem[{{Pratten} {et~al.}(2020){Pratten}, {Schmidt}, {Buscicchio}, \&
  {Thomas}}]{Pratten20}
{Pratten}, G., {Schmidt}, P., {Buscicchio}, R., \& {Thomas}, L.~M. 2020,
  \href{http://dx.doi.org/10.1103/PhysRevResearch.2.043096}{\JournalTitle{Physical
  Review Research}, 2, 043096}

\bibitem[{{Rasskazov} \& {Kocsis}(2019)}]{Rasskazov19}
{Rasskazov}, A., \& {Kocsis}, B. 2019,
  \href{http://adsabs.harvard.edu/abs/2019arXiv190203242R}{\JournalTitle{arXiv
  e-prints}}, \href{http://arxiv.org/abs/1902.03242}{{\sffamily
  arXiv:1902.03242}}

\bibitem[{{Rastello} {et~al.}(2019){Rastello}, {Amaro-Seoane}, {Arca-Sedda},
  {Capuzzo-Dolcetta}, {Fragione}, \& {Tosta e Melo}}]{Rastello18}
{Rastello}, S., {Amaro-Seoane}, P., {Arca-Sedda}, M., {et~al.} 2019,
  \href{http://dx.doi.org/10.1093/mnras/sty3193}{\JournalTitle{\mnras}, 483,
  1233}

\bibitem[{{Rastello} {et~al.}(2020){Rastello}, {Mapelli}, {Di Carlo},
  {Giacobbo}, {Santoliquido}, {Spera}, {Ballone}, \& {Iorio}}]{Rastello20}
{Rastello}, S., {Mapelli}, M., {Di Carlo}, U.~N., {et~al.} 2020,
  \href{http://dx.doi.org/10.1093/mnras/staa2018}{\JournalTitle{\mnras}, 497,
  1563}

\bibitem[{{Rodriguez} {et~al.}(2018){Rodriguez}, {Amaro-Seoane}, {Chatterjee},
  \& {Rasio}}]{Rodriguez18PRL}
{Rodriguez}, C.~L., {Amaro-Seoane}, P., {Chatterjee}, S., \& {Rasio}, F.~A.
  2018,
  \href{http://dx.doi.org/10.1103/PhysRevLett.120.151101}{\JournalTitle{\prl},
  120, 151101}

\bibitem[{{Rodriguez} {et~al.}(2016{\natexlab{a}}){Rodriguez}, {Chatterjee}, \&
  {Rasio}}]{Rodriguez16}
{Rodriguez}, C.~L., {Chatterjee}, S., \& {Rasio}, F.~A. 2016{\natexlab{a}},
  \href{http://adsabs.harvard.edu/abs/2016PhRvD..93h4029R}{\JournalTitle{Phys.
  Rev. D.}, 93, 084029}

\bibitem[{{Rodriguez} {et~al.}(2019){Rodriguez}, {Zevin}, {Amaro-Seoane},
  {Chatterjee}, {Kremer}, {Rasio}, \& {Ye}}]{Rodriguez19_hierarchical}
{Rodriguez}, C.~L., {Zevin}, M., {Amaro-Seoane}, P., {et~al.} 2019,
  \href{http://dx.doi.org/10.1103/PhysRevD.100.043027}{\JournalTitle{\prd},
  100, 043027}

\bibitem[{{Rodriguez} {et~al.}(2016{\natexlab{b}}){Rodriguez}, {Zevin},
  {Pankow}, {Kalogera}, \& {Rasio}}]{Rodriguez2016_kicks}
{Rodriguez}, C.~L., {Zevin}, M., {Pankow}, C., {Kalogera}, V., \& {Rasio},
  F.~A. 2016{\natexlab{b}},
  \href{http://dx.doi.org/10.3847/2041-8205/832/1/L2}{\JournalTitle{\apjl},
  832, L2}

\bibitem[{{Rodriguez} {et~al.}(2020){Rodriguez}, {Kremer}, {Grudi{\'c}},
  {Hafen}, {Chatterjee}, {Fragione}, {Lamberts}, {Martinez}, {Rasio},
  {Weatherford}, \& {Ye}}]{Rodriguez20_GW190412}
{Rodriguez}, C.~L., {Kremer}, K., {Grudi{\'c}}, M.~Y., {et~al.} 2020,
  \href{http://dx.doi.org/10.3847/2041-8213/ab961d}{\JournalTitle{\apjl}, 896,
  L10}

\bibitem[{{Safarzadeh} {et~al.}(2020{\natexlab{a}}){Safarzadeh}, {Farr}, \&
  {Ramirez-Ruiz}}]{Safarzadeh20}
{Safarzadeh}, M., {Farr}, W.~M., \& {Ramirez-Ruiz}, E. 2020{\natexlab{a}},
  \JournalTitle{arXiv e-prints}, arXiv:2001.06490

\bibitem[{{Safarzadeh} \& {Haiman}(2020)}]{Safarzadeh20_GW190521}
{Safarzadeh}, M., \& {Haiman}, Z. 2020, \JournalTitle{arXiv e-prints},
  arXiv:2009.09320

\bibitem[{{Safarzadeh} {et~al.}(2020{\natexlab{b}}){Safarzadeh}, {Hamers},
  {Loeb}, \& {Berger}}]{Safarzadeh20_GW190814}
{Safarzadeh}, M., {Hamers}, A.~S., {Loeb}, A., \& {Berger}, E.
  2020{\natexlab{b}},
  \href{http://dx.doi.org/10.3847/2041-8213/ab5dc8}{\JournalTitle{\apjl}, 888,
  L3}

\bibitem[{{Samsing} {et~al.}(2014){Samsing}, {MacLeod}, \&
  {Ramirez-Ruiz}}]{Samsing14}
{Samsing}, J., {MacLeod}, M., \& {Ramirez-Ruiz}, E. 2014,
  \href{http://dx.doi.org/10.1088/0004-637X/784/1/71}{\JournalTitle{\apj}, 784,
  71}

\bibitem[{{Samsing} {et~al.}(2020){Samsing}, {Bartos}, {D'Orazio}, {Haiman},
  {Kocsis}, {Leigh}, {Liu}, {Pessah}, \& {Tagawa}}]{Samsing20}
{Samsing}, J., {Bartos}, I., {D'Orazio}, D.~J., {et~al.} 2020,
  \JournalTitle{arXiv e-prints}, arXiv:2010.09765

\bibitem[{{Schmidt} {et~al.}(2015){Schmidt}, {Ohme}, \& {Hannam}}]{Schmidt2015}
{Schmidt}, P., {Ohme}, F., \& {Hannam}, M. 2015,
  \href{http://dx.doi.org/10.1103/PhysRevD.91.024043}{\JournalTitle{\prd}, 91,
  024043}

\bibitem[{{Sch{\"o}del} {et~al.}(2020){Sch{\"o}del}, {Nogueras-Lara},
  {Gallego-Cano}, {Shahzamanian}, {Gallego-Calvente}, \& {Gardini}}]{Schodel20}
{Sch{\"o}del}, R., {Nogueras-Lara}, F., {Gallego-Cano}, E., {et~al.} 2020,
  \JournalTitle{arXiv e-prints}, arXiv:2007.15950

\bibitem[{Scott(1992)}]{Scott92}
Scott, D. 1992, Multivariate Density Estimation: Theory, Practice, and
  Visualization, A Wiley-interscience publication (Wiley)

\bibitem[{{Silsbee} \& {Tremaine}(2017)}]{Silsbee17}
{Silsbee}, K., \& {Tremaine}, S. 2017,
  \href{http://adsabs.harvard.edu/abs/2017ApJ...836...39S}{\JournalTitle{\apj},
  836, 39}

\bibitem[{{Spera} {et~al.}(2019){Spera}, {Mapelli}, {Giacobbo}, {Trani},
  {Bressan}, \& {Costa}}]{Spera19}
{Spera}, M., {Mapelli}, M., {Giacobbo}, N., {et~al.} 2019,
  \href{http://dx.doi.org/10.1093/mnras/stz359}{\JournalTitle{\mnras}, 485,
  889}

\bibitem[{{Stone} {et~al.}(2017){Stone}, {Metzger}, \& {Haiman}}]{Stone17}
{Stone}, N.~C., {Metzger}, B.~D., \& {Haiman}, Z. 2017,
  \href{http://adsabs.harvard.edu/abs/2017MNRAS.464..946S}{\JournalTitle{\mnras},
  464, 946}

\bibitem[{{Tagawa} {et~al.}(2020{\natexlab{a}}){Tagawa}, {Haiman}, {Bartos}, \&
  {Kocsis}}]{Tagawa20b_spin}
{Tagawa}, H., {Haiman}, Z., {Bartos}, I., \& {Kocsis}, B. 2020{\natexlab{a}},
  \href{http://dx.doi.org/10.3847/1538-4357/aba2cc}{\JournalTitle{\apj}, 899,
  26}

\bibitem[{{Tagawa} {et~al.}(2020{\natexlab{b}}){Tagawa}, {Haiman}, \&
  {Kocsis}}]{Tagawa19}
{Tagawa}, H., {Haiman}, Z., \& {Kocsis}, B. 2020{\natexlab{b}},
  \href{http://dx.doi.org/10.3847/1538-4357/ab9b8c}{\JournalTitle{\apj}, 898,
  25}

\bibitem[{{Tagawa} {et~al.}(2021{\natexlab{a}}){Tagawa}, {Kocsis}, {Haiman},
  {Bartos}, {Omukai}, \& {Samsing}}]{Tagawa20_ecc}
{Tagawa}, H., {Kocsis}, B., {Haiman}, Z., {et~al.} 2021{\natexlab{a}},
  \href{http://dx.doi.org/10.3847/2041-8213/abd4d3}{\JournalTitle{\apjl}, 907,
  L20}

\bibitem[{{Tagawa} {et~al.}(2021{\natexlab{b}}){Tagawa}, {Kocsis}, {Haiman},
  {Bartos}, {Omukai}, \& {Samsing}}]{Tagawa20_massgap}
---. 2021{\natexlab{b}},
  \href{http://dx.doi.org/10.3847/1538-4357/abd555}{\JournalTitle{\apj}, 908,
  194}

\bibitem[{{Tagawa} {et~al.}(2018){Tagawa}, {Kocsis}, \& {Saitoh}}]{Tagawa18}
{Tagawa}, H., {Kocsis}, B., \& {Saitoh}, R.~T. 2018,
  \href{http://adsabs.harvard.edu/abs/2018PhRvL.120z1101T}{\JournalTitle{Phys.
  Rev. Lett.}, 120, 261101}

\bibitem[{{The LIGO Scientific Collaboration} {et~al.}(2019){The LIGO
  Scientific Collaboration}, {the Virgo Collaboration}, {Abbott}, {Abbott},
  {Abbott}, {Abraham}, {Acernese}, {Ackley}, {Adams}, {Adams}, {Adhikari},
  {Adya}, {Affeldt}, {Agathos}, {Agatsuma}, {Aggarwal}, {Aguiar}, {Aiello},
  {Ain}, {Ajith}, {Allen}, {Allocca}, {Aloy}, {Altin}, {Amato}, {Anand},
  {Ananyeva}, {Anderson}, {Anderson}, {Angelova}, \& {Antier}}]{LIGO19_IMBH}
{The LIGO Scientific Collaboration}, {the Virgo Collaboration}, {Abbott},
  B.~P., {et~al.} 2019, \JournalTitle{arXiv e-prints}, arXiv:1906.08000

\bibitem[{{The LIGO Scientific Collaboration} {et~al.}(2020{\natexlab{a}}){The
  LIGO Scientific Collaboration}, {the Virgo Collaboration}, {Abbott},
  {Abbott}, {Abraham}, {Acernese}, {Ackley}, {Adams}, {Adhikari}, {Adya},
  {Affeldt}, {Agathos}, {Agatsuma}, {Aggarwal}, {Aguiar}, {Aich}, {Aiello},
  {Ain}, \& {Ajith}}]{LIGO20_GW190412}
{The LIGO Scientific Collaboration}, {the Virgo Collaboration}, {Abbott}, R.,
  {et~al.} 2020{\natexlab{a}}, \JournalTitle{arXiv e-prints}, arXiv:2004.08342

\bibitem[{{The LIGO Scientific Collaboration} {et~al.}(2020{\natexlab{b}}){The
  LIGO Scientific Collaboration}, {the Virgo Collaboration}, {Abbott},
  {Abbott}, {Abraham}, {Acernese}, {Ackley}, {Adams}, {Adhikari}, {Adya},
  {Affeldt}, {Agathos}, {Agatsuma}, {Aggarwal}, {Aguiar}, {Aich}, {Aiello},
  {Ain}, {Ajith}, {Akcay}, {Allen}, {Allocca}, {Altin}, {Amato}, {Anand},
  {Ananyeva}, {Anderson}, {Anderson}, {Angelova}, \&
  {Ansoldi}}]{LIGO20_GW190521}
---. 2020{\natexlab{b}}, \JournalTitle{arXiv e-prints}, arXiv:2009.01075

\bibitem[{{The LIGO Scientific Collaboration} {et~al.}(2020{\natexlab{c}}){The
  LIGO Scientific Collaboration}, {the Virgo Collaboration}, {Abbott},
  {Abbott}, {Abraham}, {Acernese}, {Ackley}, {Adams}, {Adams}, \&
  {Adhikari}}]{LIGO20_O3_Properties}
---. 2020{\natexlab{c}}, \JournalTitle{arXiv e-prints}, arXiv:2010.14533

\bibitem[{{Tiwari} \& {Fairhurst}(2020)}]{Tiwari20}
{Tiwari}, V., \& {Fairhurst}, S. 2020, \JournalTitle{arXiv e-prints},
  arXiv:2011.04502

\bibitem[{{van den Heuvel} {et~al.}(2017){van den Heuvel}, {Portegies Zwart},
  \& {de Mink}}]{vandenHeuvel17}
{van den Heuvel}, E.~P.~J., {Portegies Zwart}, S.~F., \& {de Mink}, S.~E. 2017,
  \href{http://adsabs.harvard.edu/abs/2017MNRAS.471.4256V}{\JournalTitle{\mnras},
  471, 4256}

\bibitem[{{Veitch} {et~al.}(2015){Veitch}, {Raymond}, {Farr}, {Farr}, {Graff},
  {Vitale}, {Aylott}, {Blackburn}, {Christensen}, {Coughlin}, {Del Pozzo},
  {Feroz}, {Gair}, {Haster}, {Kalogera}, {Littenberg}, {Mandel},
  {O'Shaughnessy}, {Pitkin}, {Rodriguez}, {R{\"o}ver}, {Sidery}, {Smith}, {Van
  Der Sluys}, {Vecchio}, {Vousden}, \& {Wade}}]{Veitch15}
{Veitch}, J., {Raymond}, V., {Farr}, B., {et~al.} 2015,
  \href{http://dx.doi.org/10.1103/PhysRevD.91.042003}{\JournalTitle{\prd}, 91,
  042003}

\bibitem[{{Venumadhav} {et~al.}(2019){Venumadhav}, {Zackay}, {Roulet}, {Dai},
  \& {Zaldarriaga}}]{Venumadhav19}
{Venumadhav}, T., {Zackay}, B., {Roulet}, J., {Dai}, L., \& {Zaldarriaga}, M.
  2019,
  \href{http://adsabs.harvard.edu/abs/2019arXiv190407214V}{\JournalTitle{arXiv
  e-prints}}, \href{http://arxiv.org/abs/1904.07214}{{\sffamily
  arXiv:1904.07214}}

\bibitem[{{Vink} {et~al.}(2021){Vink}, {Higgins}, {Sander}, \&
  {Sabhahit}}]{Vink2021}
{Vink}, J.~S., {Higgins}, E.~R., {Sander}, A. A.~C., \& {Sabhahit}, G.~N. 2021,
  \href{http://dx.doi.org/10.1093/mnras/stab842}{\JournalTitle{\mnras}, 504,
  146}

\bibitem[{{Vitale} {et~al.}(2020){Vitale}, {Gerosa}, {Farr}, \&
  {Taylor}}]{2020arXiv200705579V}
{Vitale}, S., {Gerosa}, D., {Farr}, W.~M., \& {Taylor}, S.~R. 2020,
  \JournalTitle{arXiv e-prints}, arXiv:2007.05579

\bibitem[{{Yang} {et~al.}(2020{\natexlab{a}}){Yang}, {Bartos}, {Haiman},
  {Kocsis}, {M{\'a}rka}, \& {Tagawa}}]{Yang20}
{Yang}, Y., {Bartos}, I., {Haiman}, Z., {et~al.} 2020{\natexlab{a}},
  \JournalTitle{arXiv e-prints}, arXiv:2003.08564

\bibitem[{{Yang} {et~al.}(2020{\natexlab{b}}){Yang}, {Gayathri}, {Bartos},
  {Haiman}, {Safarzadeh}, \& {Tagawa}}]{Yang20_GW190814}
{Yang}, Y., {Gayathri}, V., {Bartos}, I., {et~al.} 2020{\natexlab{b}},
  \JournalTitle{arXiv e-prints}, arXiv:2007.04781

\bibitem[{{Yang} {et~al.}(2019){Yang}, {Bartos}, {Gayathri}, {Ford}, {Haiman},
  {Klimenko}, {Kocsis}, {M{\'a}rka}, {M{\'a}rka}, {McKernan}, \&
  {O'Shaughnessy}}]{Yang19b_PRL}
{Yang}, Y., {Bartos}, I., {Gayathri}, V., {et~al.} 2019,
  \href{http://dx.doi.org/10.1103/PhysRevLett.123.181101}{\JournalTitle{\prl},
  123, 181101}

\bibitem[{{Yoon} {et~al.}(2012){Yoon}, {Dierks}, \& {Langer}}]{Yoon12}
{Yoon}, S.~C., {Dierks}, A., \& {Langer}, N. 2012,
  \href{http://dx.doi.org/10.1051/0004-6361/201117769}{\JournalTitle{\aap},
  542, A113}

\bibitem[{{Zackay} {et~al.}(2019){Zackay}, {Dai}, {Venumadhav}, {Roulet}, \&
  {Zaldarriaga}}]{Zackay19_GW170817A}
{Zackay}, B., {Dai}, L., {Venumadhav}, T., {Roulet}, J., \& {Zaldarriaga}, M.
  2019, \JournalTitle{arXiv e-prints}, arXiv:1910.09528

\bibitem[{{Zevin} {et~al.}(2020{\natexlab{a}}){Zevin}, {Spera}, {Berry}, \&
  {Kalogera}}]{Zevin2020_GW190814}
{Zevin}, M., {Spera}, M., {Berry}, C. P.~L., \& {Kalogera}, V.
  2020{\natexlab{a}},
  \href{http://dx.doi.org/10.3847/2041-8213/aba74e}{\JournalTitle{\apjl}, 899,
  L1}

\bibitem[{{Zevin} {et~al.}(2020{\natexlab{b}}){Zevin}, {Bavera}, {Berry},
  {Kalogera}, {Fragos}, {Marchant}, {Rodriguez}, {Antonini}, {Holz}, \&
  {Pankow}}]{Zevin2020_multiple}
{Zevin}, M., {Bavera}, S.~S., {Berry}, C. P.~L., {et~al.} 2020{\natexlab{b}},
  \JournalTitle{arXiv e-prints}, arXiv:2011.10057

\bibitem[{{Ziosi} {et~al.}(2014){Ziosi}, {Mapelli}, {Branchesi}, \&
  {Tormen}}]{Ziosi14}
{Ziosi}, B.~M., {Mapelli}, M., {Branchesi}, M., \& {Tormen}, G. 2014,
  \href{http://dx.doi.org/10.1093/mnras/stu824}{\JournalTitle{\mnras}, 441,
  3703}

\end{thebibliography}

\appendix

\begin{figure*}
\begin{center}
\includegraphics[width=180mm]{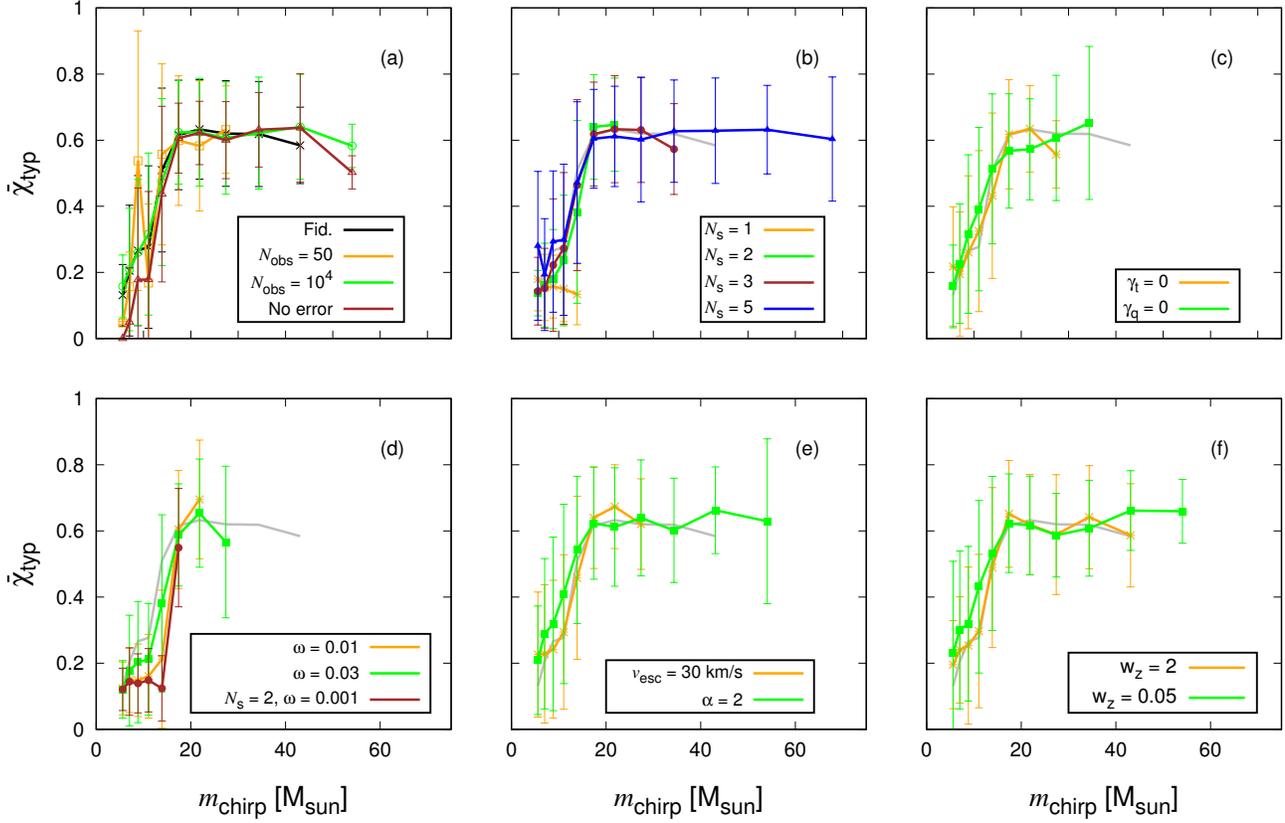}
\caption{
Same as Fig.~\ref{fig:x_mc_profile_dep}, but present 
for models~M1, M14--M28 (Table~\ref{table:fraction_highg}). 
We use $N_\mathrm{obs}=10^3$ detectable mergers for models~M1, M14--M28, while $N_\mathrm{obs}=50$ and $10^4$ for models~M14 and M15, respectively. 
}
\label{fig_app:x_mc_profile_dep}
\end{center}
\end{figure*}

\begin{table*}
\begin{center}
\caption{
Same as Table~\ref{table:fraction_highg}, but includes models~M14--M28. 
}
\label{table_app:fraction_highg}
\hspace{-5mm}
\begin{tabular}{c|c|c|c|c|c|c}
\hline 
model&Parameter & high-g fraction & high-g detection fraction 
& $m_\mathrm{chirp,max} [\Msun]$ & ${\bar \chi}_\mathrm{p}$&$\sigma(\chi_\mathrm{p})$
\\\hline\hline
M1&Fiducial & 0.33 &0.68&56&0.17 & 0.26\\\hline
M2&Globular cluster (GC) & 0.063 & 0.17 & 44&0.030&0.13\\\hline
M3&Field binary (FB) & 0 & 0 & 23&0&0\\\hline
M4&Migration trap (MT) & 0.31 & 0.80 & 42&0&0\\\hline
M5&$a_\mathrm{uni}=1$ & 0.32 & 0.73 & 52&0.50&0.21\\\hline
M6&$a_\mathrm{ave}=0.99$ & 0.31 & 0.70 & 51&0.75&0.20\\\hline
M7&$a_\mathrm{ave}=0.66$, $a_\mathrm{uni}=0.1$ & 0.33 & 0.72 & 55&0.55&0.13\\\hline
M8&$a_\mathrm{ave}=0.5$ & 0.33 & 0.74 & 65&0.46&0.12\\\hline
M9&$m_\mathrm{max}=30\,\Msun$ & 0.35 & 0.73 & 70&0.18&0.26\\\hline
M10&$N_\mathrm{obs}=50$, $N_\mathrm{s}=3$& 0.25& 0.62 & 28&0.13&0.24\\\hline
M11&$N_\mathrm{obs}=50$, $N_\mathrm{s}=2$& 0.15& 0.28 & 24&0.077&0.19\\\hline
M12&$N_\mathrm{obs}=50$, $N_\mathrm{s}=2$, $\omega=0.05$& 0.077& 0.18 & 19&0.040&0.14\\\hline
M13&$N_\mathrm{obs}=50$, $N_\mathrm{s}=2$, $\omega=0.03$& 0.046& 0.14 & 19&0.023&0.11\\\hline
M14&$N_\mathrm{obs}=50$ & 0.33 & 0.78 & 38&0.17&0.26\\\hline
M15&$N_\mathrm{obs}=10^4$ & 0.33 & 0.77 & 60&0.17&0.26\\\hline
M16&$N_\mathrm{s}=1$ & 0 & 0 & 15&0&0\\\hline
M17&$N_\mathrm{s}=2$& 0.15& 0.40 & 23&0.077&0.19\\\hline
M18&$N_\mathrm{s}=3$& 0.25& 0.61 & 42&0.13&0.24\\\hline
M19&$N_\mathrm{s}=5$& 0.38& 0.80 & 72&0.20&0.27\\\hline
M20&$\gamma_\mathrm{t}=0$ & 0.22 & 0.50 & 31&0.11&0.22\\\hline
M21&$\gamma_\mathrm{q}=0$ & 0.39 & 0.71 & 36&0.19&0.25\\\hline
M22&$\omega=0.01$ & 0.043 & 0.089 & 23&0.022&0.11\\\hline
M23&$\omega=0.03$ & 0.13 & 0.37 & 33&0.066&0.18\\\hline
M24&$N_\mathrm{s}=2$, $\omega=0.001$& 0.0014& 0.0040 & 17&0.001&0.02\\\hline
M25&$v_\mathrm{esc}=30\,\mathrm{km/s}$ & 0.29 & 0.61 &31&0.15&0.25\\\hline
M26&$\alpha=2$ & 0.32 & 0.75 & 52&0.16&0.25\\\hline
M27&$w_\mathrm{z}=2$ & 0.33 & 0.73 & 46&0.17&0.26\\\hline
M28&$w_\mathrm{z}=0.05$ & 0.33 & 0.79 & 59&0.17&0.26\\\hline
\end{tabular}
\end{center}
\end{table*}

\section{Dependence on population parameters}

\label{sec_app:result_pop_parameters}

We show the parameter dependence of the ${\bar \chi}_\mathrm{typ}$ profile as a function of $m_\mathrm{chirp}$ using mock GW events, in which hierarchical mergers are assumed to be frequent. 
In Table~\ref{table_app:fraction_highg}, we list the model varieties we have investigated (models~M1--M28). 
We additionally examine different choices of 
the number of detected mergers (models~M14 and M15), 
the steps to create samples for hierarchical mergers (models~M16--M19), 
pairing probability (models~M20 and M21), 
the fraction of mergers in each step (models~M22--M24), 
the escape velocity of the system (model~M25), 
the power law for mass function (model~M26), 
and the correlation between the steps and the redshift (models~M27 and M28).

With smaller number of iteration steps ($N_\mathrm{s}$), the maximum $m_\mathrm{chirp}$ becomes smaller because the generations of BHs are limited by $N_\mathrm{s}$ (panel~b of Fig.~\ref{fig_app:x_mc_profile_dep}). 
Similarly, the maximum $m_\mathrm{chirp}$ decreases as 
$N_\mathrm{obs}$, 
$\gamma_\mathrm{t}$, $\omega$, or $v_\mathrm{esc}$ decreases or $m_\mathrm{max}$ increases (panels~a, c, d, and e of Fig.~\ref{fig_app:x_mc_profile_dep} and panel~b of Fig.~\ref{fig:x_mc_profile_dep}, Table~\ref{table_app:fraction_highg}). 
In these ways, the maximum $m_\mathrm{chirp}$ is influenced by a number of parameters, implying that the maximum $m_\mathrm{chirp}$ alone cannot constrain each of those parameters.

Here, we investigate the effect that mergers at larger iteration steps tend to occur at lower redshift because finite time needs to elapse between each generation and 
high-g mergers thus would take place after a significant delay compared to low-g mergers.
To take this delay into account, we modify the redshift distribution of merging BHs as 
\begin{align}
\label{eq:p_z_ev}
   p_{z}\propto &\frac{dV_c}{dz}\frac{1}{1+z}\mathrm{exp}\left(\frac{(t_\mathrm{L}(z)-\mu_t)^2}{2\sigma_t^2}\right),
\end{align}
where $t_\mathrm{L}(z)$ is the look-back time, 
we set the average to  $\mu_t=t_\mathrm{typ}\left(\frac{N_\mathrm{s}-N_i +1}{ N_\mathrm{s}}\right)$ and the standard deviation to  $\sigma_t=t_\mathrm{typ} w_t$, $N_i$ is the number of steps that the $i^\mathrm{th}$ merger is created, $t_\mathrm{typ}$ is the typical look-back time that mergers began to occur, which is set to $10\,\mathrm{Gyr}$, and $w_z$ is the parameter determining the strength of correlation between $N_i$ and the time that mergers occur. A lower value of $w_z$ makes mergers with high $N_i$ occur at a lower $z$, and the fiducial model (Eq.~\ref{eq:p_z}) corresponds to $w_z=\infty$. 
The dependence of the ${\bar \chi}_\mathrm{typ}$ profile on $w_z$ 
is shown in panel~(f), suggesting that the correlation between the redshift and the generations of BHs has a negligible impact on the profile.

\section{Observed spin distribution}
\label{sec_app:spin_dist}

\begin{figure*}
\begin{center}
\includegraphics[width=180mm]{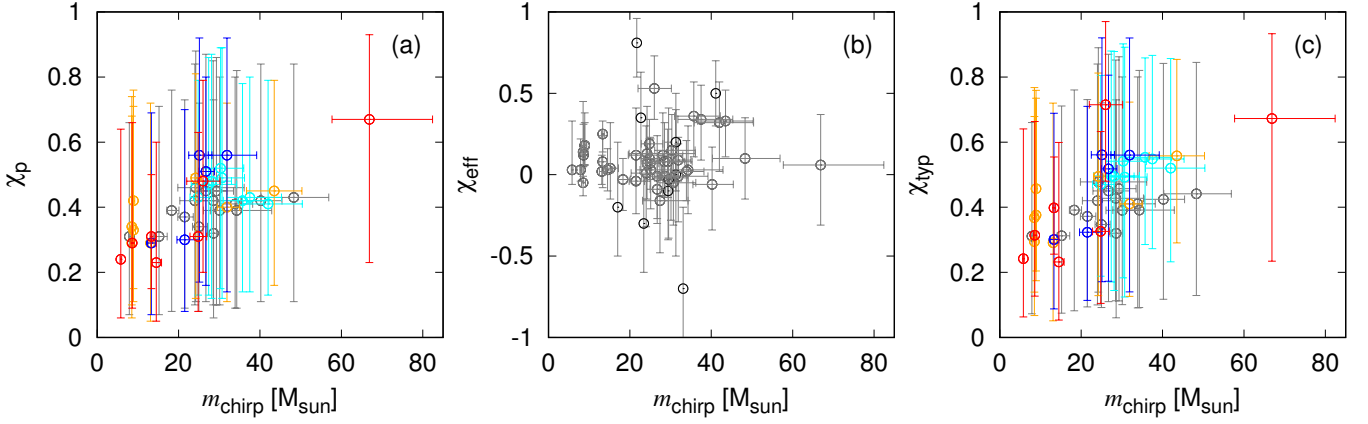}
\caption{
Observed distribution for spins reported in LIGO/Virgo O1--O3a. Panels~(a)--(c) represent the distributions of $\chi_\mathrm{p}$, $\chi_\mathrm{eff}$, and $\chi_\mathrm{typ}$, respectively. 
Red, blue, orange, cyan, and gray circles represent events with the KL divergence between prior and posterior samples for $\chi_\mathrm{p}$ to be  $D_\mathrm{KL}\geq0.2$, 
$0.2>D_\mathrm{KL}\geq0.15$, 
$0.15>D_\mathrm{KL}\geq0.1$, 
$0.1>D_\mathrm{KL}\geq0.05$, and 
$0.05>D_\mathrm{KL}$, respectively. 
Bars correspond to the $90$ percentile credible intervals. 
}
\label{fig_app:mc_x_obs}
\end{center}
\end{figure*}

\begin{figure}
\begin{center}
\includegraphics[width=95mm]{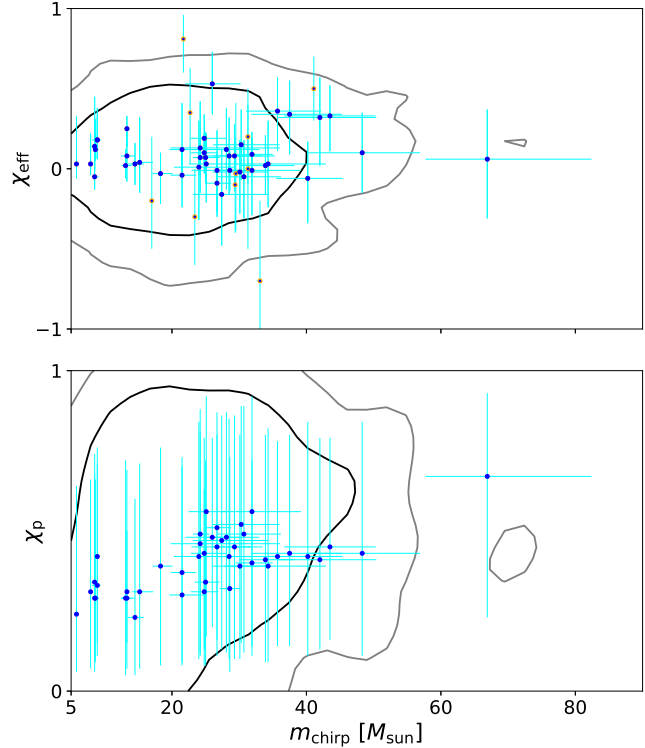}
\caption{
Comparisons between the $\chi_\mathrm{eff}$ (upper) or $\chi_\mathrm{p}$ (lower) and $m_\mathrm{chirp}$ distributions observed by LIGO/Virgo O1--O3a and those predicted by the model with $m_\mathrm{max}=30\,\Msun$ and $\alpha=2$ which have high Bayes factors (Table~\ref{table:bayesian_values}). 
Black and gray lines correspond to 90 and 99 credible intervals for the predicted distributions, and cyan bars correspond to the 90 percentile credible intervals for the observed variables. 
The orange points in the upper panel corresponds to the events reported in \citet{Venumadhav19} and \citet{Zackay19_GW170817A}. 
}
\label{fig_app:mc_x_obs9}
\end{center}
\end{figure}

We presents the observed distributions of $\chi_\mathrm{p}$, $\chi_\mathrm{eff}$, and $\chi_\mathrm{typ}$ as a function of $m_\mathrm{chirp}$ in Fig.~\ref{fig_app:mc_x_obs}. 
Also, Fig.~\ref{fig_app:mc_x_obs9} compares the $m_\mathrm{chirp}$, $\chi_\mathrm{eff}$ and $\chi_\mathrm{p}$ distributions observed by LIGO/Virgo O1--O3a and those predicted by the model for the fiducial settings (Table~\ref{table:parameter_fiducial}) but $m_\mathrm{max}=30\,\Msun$ and $\alpha=2$, which is assessed to high Bayes factors for both $D_\mathrm{KL,crit}=0.05$ and the two parameters (Table~\ref{table:bayesian_values}). 
We can see that the observed distribution for these variables (blue and orange points) roughly follows the 90 and 99 percentile regions (black and gray lines) predicted by the model.

\section{Posterior distributions for spin parameters}

\begin{figure*}
\begin{center}
\includegraphics[width=170mm]{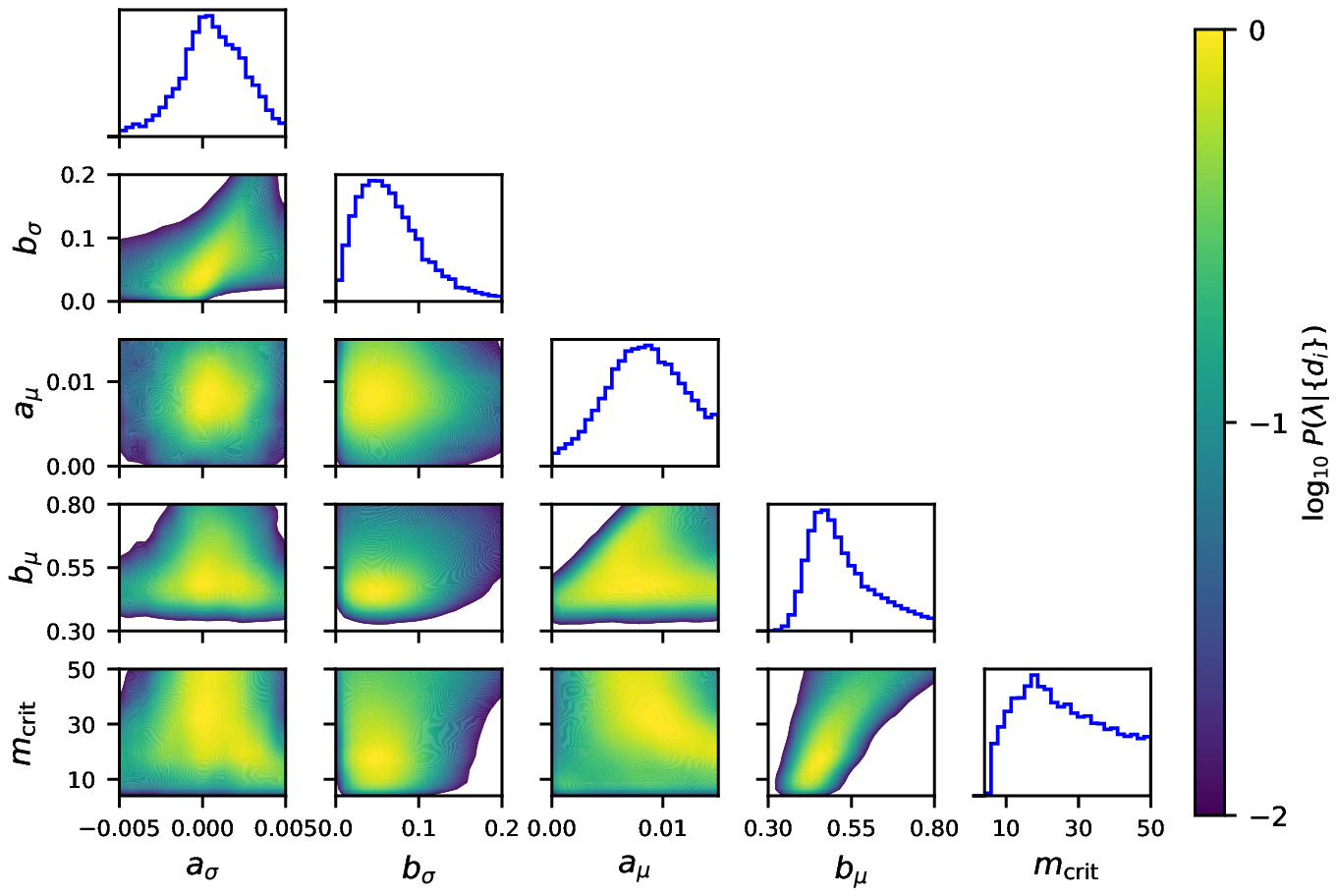}
\caption{
The posterior probability distributions for $a_\sigma$, $b_\sigma$, $a_\mu$, $b_\mu$, and $m_{\rm crit}$ for $D_{\rm KL,crit}=0$ ($\S\,\ref{sec:result_rec_x_obs}$). 
}
\label{fig_app:mc_dist}
\end{center}
\end{figure*}

We present the posterior distributions of the parameters characterizing the spin profile for the GW events with $D_{\rm KL}\geq0$ ($\S\,\ref{sec:result_rec_x_obs}$) in Fig.~\ref{fig_app:mc_dist}.

\end{document}